%% file: Submitted_onecolumn.tex
\documentclass[onecolumn,12pt,draftclsnofoot]{IEEEtran}
\IEEEoverridecommandlockouts

\usepackage{cite}

\usepackage{graphicx}
\usepackage[cmex10]{amsmath}
\usepackage{amsfonts,amssymb,bm}
\usepackage[isolatin]{inputenc}
\usepackage{array}
\usepackage{multirow}
\usepackage{subcaption} %it includes also the subfig package
\usepackage{color}

\DeclareMathOperator{\tr}{tr}

\newcommand{\smse}{\mathop{\rm SMSE}}
\input{user_cmds}

\begin{document}
\title{Distributed Precoding Systems in Multi-Gateway  Multibeam Satellites: Regularization and Coarse Beamforming}

\author{Carlos Mosquera,~\IEEEmembership{Senior~Member,~IEEE}, Roberto L\'opez-Valcarce,~\IEEEmembership{Member,~IEEE}, 
Vahid Joroughi \thanks{
Carlos Mosquera and Roberto L\'opez-Valcarce are with the Signal Theory and Communications Department, University of Vigo, Galicia, Spain. (e-mail: \{mosquera, valcarce\}@gts.uvigo.es). Vahid Joroughi was with the  Signal Theory and Communications Department, University of Vigo, Galicia, Spain.
This work was partially funded by the Agencia Estatal de Investigaci\'on (Spain) and the European Regional Development Fund (ERDF) through the
Projects COMPASS under Grant TEC2013-47020-C2-1-R, MYRADA under
Grant TEC2016-75103-C2-2-R, WINTER under Grant TEC2016-76409-C2-
2-R, and COMONSENS under Grant TEC2015-69648-REDC.
Also funded by the Xunta de Galicia (Agrupaci\'on Estrat\'exica Consolidada de Galicia accreditation
2016-2019; Red Temática RedTEIC 2017-2018) and the European Union (European Regional Development Fund - ERDF).
}}

\maketitle

\begin{abstract}

This paper deals with the problem of beamforming design in a multibeam satellite, which is shared by different groups of terminals -clusters-,
 each served by an Earth  station or gateway. Each gateway precodes the symbols addressed to its respective  users;
 the design  follows an MMSE 
criterion, and a regularization factor judiciously chosen allows to account for the presence of mutually interfering clusters, extending more classical results applicable to one centralized station. More importantly, channel statistics can be used instead of instantaneous channel state 
information, avoiding the exchange of information among gateways through backhaul links. 
The on-board satellite beamforming 
weights are designed to exploit the degrees of freedom of the satellite antennas to minimize the noise impact and the interference to some specific users. On-ground beamforming results are provided as a reference to compare the joint performance of MMSE precoders and on-board 
beamforming network.
A non-adaptive design complements the results and makes them more amenable to practical use by designing a coarse beamforming network. 

\end{abstract}

%\begin{IEEEkeywords}
% MIMO satellite, multipe gateway systems, multibeam satellite systems, on-board beamforming.
%\end{IEEEkeywords}

%\IEEEpeerreviewmaketitle

\section{Introduction}

Relaying is a widespread mechanism to extend the coverage of wireless links by appropriate manipulation of the input signals, which might include amplification, filtering and  beamforming, just to mention a few  physical layer operations.  
Bent-pipe communication satellites can be considered as non-regenerative relays \cite{Minoli2015}, essentially filtering and amplifying signals, 
although they are very complex communication systems and handle simultaneously many streams of information. The object of this study is a multibeam satellite which relays the signals coming 
from $M$ ground stations (gateways)  to convey  their communication with single-antenna terminals. 
The  foot-print of a multibeam satellite is made of many spot-beams, hundreds in some specific commercial cases, 
which are synthesized by  the on-board beamforming network (BFN) in combination with the antennas radiation pattern. Two implementation 
approaches are possible: single feed per beam and multiple feeds per beam. For the purpose of this paper, 
it is of specific interest the case of multiple  feeds per beam, for which small subarrays are used for each spot-beam\footnote{We will use spot-beam and beam as equivalent terms in this paper.}, and adjacent 
spot-beams share some of the array elements. This technology has some advantages since individual beams can overlap and a single reflector antenna served 
by several feeds can cover a larger area \cite{Minoli2015}. The  on-board beamforming 
(OBBF) process contributes some flexibility to the shaping of the beams, although the configurability is in most cases quite limited, and real-time 
adaptation in the range of milliseconds is rarely feasible. Remarkably,  a technology known as On-Ground Beamforming (OGBF) 
has emerged as an alternative 
solution for some specific cases, to avoid the need for a complex on-board digital processor. 
This technology has been used in some recent 
multibeam mobile satellite systems \cite{Tronc2014}, 
and requires the exchange  of all the feed signals between the satellite and the gateway, 
increasing the bandwidth demands on the feeder link due to the higher number of information streams. 

\begin{figure*}
\centering
\includegraphics[width=0.9\textwidth]{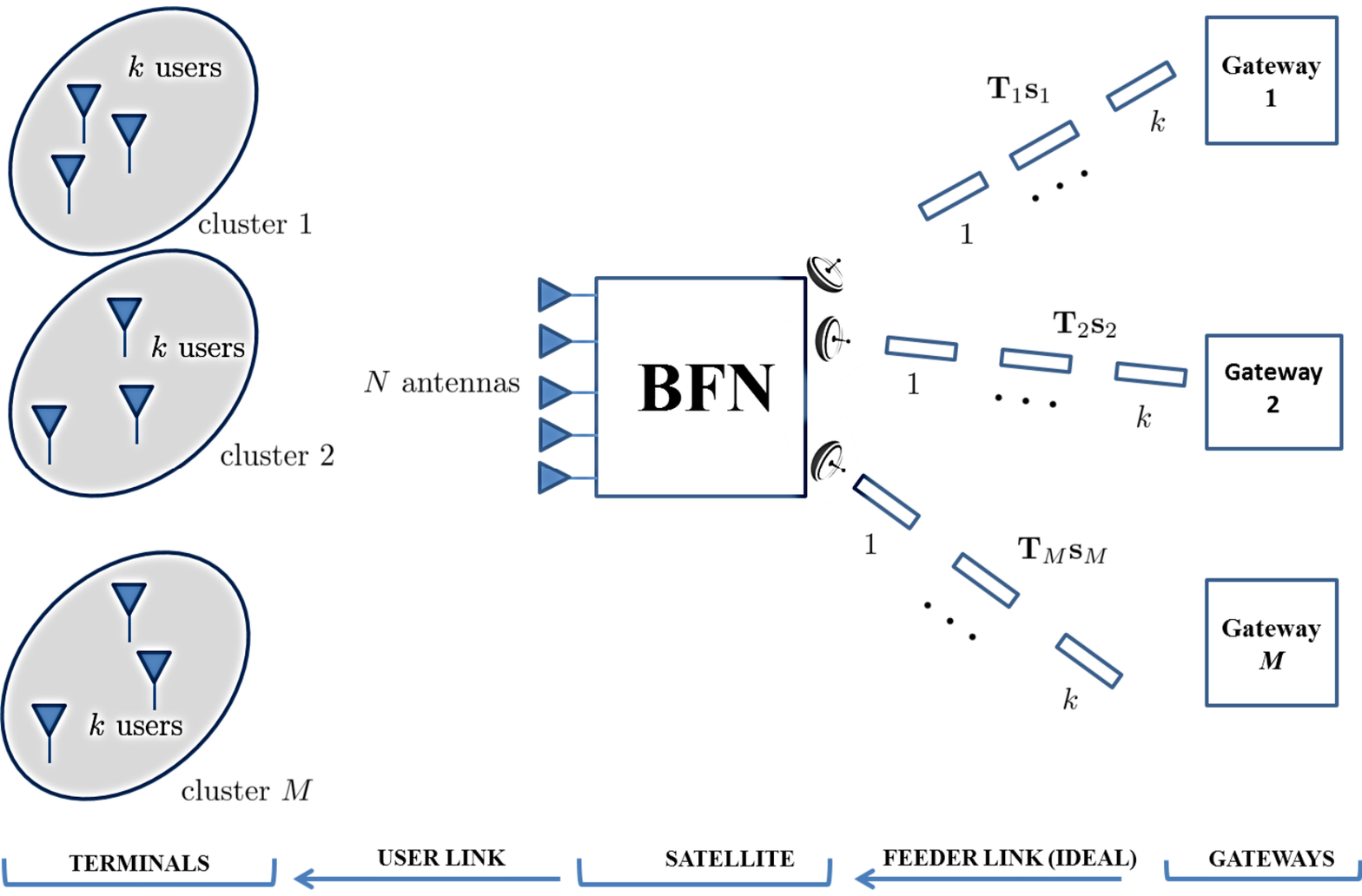}
\caption{Satellite shared by a number of ground stations.}
\label{fig:system_des}
\end{figure*}

\begin{figure}
\centering
\includegraphics[width=0.5\textwidth]{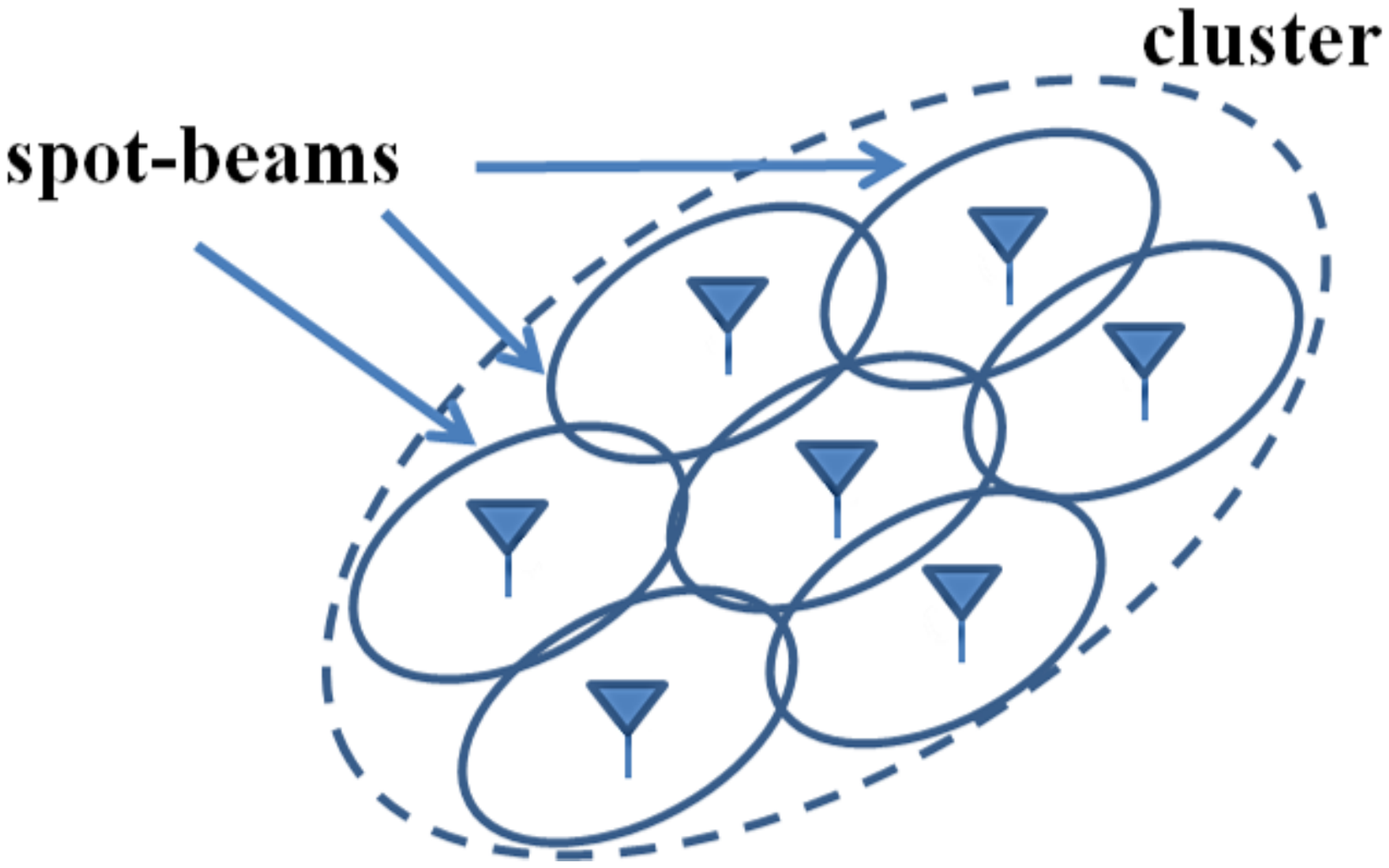}
\caption{Cluster of beams.}
\label{fig:cluster}
\end{figure}

Figure \ref{fig:system_des} depicts the conceptual abstraction  of the multibeam satellite operation, with the following features to be highlighted: 
(i) the feeder links, from the gateways to the satellites, can be assumed transparent, whereas the user links are frequency non-selective; 
(ii) there is no interference between feeder links and user links, since the communication 
 takes place on  different frequency bands; 
(iii) a given cluster is made of several beams (see Figure \ref{fig:cluster}),  with one user per beam served at a time by 
 a given frequency carrier;  (iv) the user link frequency carriers are made available to all beams and clusters, in what it is known as 
full-frequency reuse. 

One  major challenge for multibeam satellites 
is the large spectral demand on the feeder link between the satellite and the operator 
stations, since it has to aggregate the traffic from all beams. 
Technology has contributed to a steady increase of this traffic during the last years due to, 
among other things,  a more efficient reuse of spectrum across the different beams \cite{Zheng2012}.  The use of different gateways can generate several parallel channels provided that the antennas guarantee the required spatial isolation, which is the case for frequencies in  Ku-band  or
Ka-band. Thus, the different feeder links can reuse the whole available bandwidth. \\

The preprocessing of signals to communicate multiple-antennas in one-site with many users simultaneously is supported by theoretical bounds and practical schemes presented in many references. In the particular case of linear precoding,  the seminal paper \cite{Hochwald2005a} analyzes the regularization of channel inversion at the transmitter to maximize the signal to interference and noise ratio (SINR), with specific focus on Rayleigh channels. On the other side, precoding for multibeam satellites has been extensively explored in the literature to fight interbeam interference 
in the case of a single gateway, see, 
e.g., \cite{DevillersNeiraMosquera} and \cite{GaudenziBook} among others. Potential gain and calibration needs have been properly identified 
by these works. 
As opposed, results for multiple gateways are still incomplete; a centralized multi-gateway resource management, 
which for mathematical purposes can be assumed, is far from being realistic in practice \cite{Zheng2012}, due to the backhaul links that would 
be needed to connect all the gateways. Some precoding schemes  for multiple  gateways without  BFN are presented in  \cite{Vahid2016}, which 
assume  the exchange of information for the design of their respective precoders. 
However, no integral 
approach conciliating the design of an on-board BFN and the use of several gateways, required to channelize the high throughput for an aggressive use of the user link spectrum, is known. 
The mapping of groups of beams to different gateways prevents from a centralized management, 
with inter-cluster interference more difficult to control.\\

 In this work we  try to keep the cooperation at a minimum, so that no 
 information symbols are exchanged among the terrestrial gateways, each communicating with the terminals operating on its cluster. 
Initially each gateway is expected to know the channel state information (CSI) of the links originating from itself, including inter-cluster 
links, although the use of channel statistics is shown to predict  quite conveniently the required information so 
the interaction among different clusters can be completely avoided. The global interplay of distributed precoders operating 
 at the different gateways and the BFN on the satellite can be designed in such a way that different solutions accrue, each fitting the 
available information and flexibility of the involved subsystems.  
Specific regularization rules for the distributed precoders are  obtained as an 
extension of the results for one cluster.   
As a reference, 
we  compute the performance of the fully flexible OGBF with perfect CSI at both the gateways (CSIT) and receive terminals (CSIR).
  Starting from that, separate on-board BFN and precoders are derived. 
The performance indicator is the Mean Square Error (MSE) for three different settings: 
(i) intra-cluster interference driven; (ii) cancellation of interference leaked to specific off-cluster users; 
(iii) coarse fixed BFN designed as a trade-off solution.
The latter scheme is particularly important, since it allows to fix the on-board BFN and confine the flexibility to  the gateway precoders: 
some loss of performance is expected to achieve this more practical design. \\

After detailing the satellite relaying  operation in Section \ref{sec:relay}, we  derive the optimal beamforming weights and 
gateway precoders, first for the OGBF case in Section \ref{sec:OGBF} and then for separate on-board beamforming and ground precoders in Section 
\ref{sec:OBBF}. Both fixed and adaptive BFN weights are object of the study, with performance tested in the simulations shown before the conclusions. \\

%Insist: cooperative techniques, limited feedback, partial cooperation, maximize sum-rate at high SNR

\noindent {\textit Notation:} Upper (lower) boldface letters denote matrices 
(vectors). $(.)^H$, $(.)^T $, $\trace\{\cdot\}$, $\mathbf{I}_N$, $\bm 0$, 
$\text{diag}\{\cdot \}$   denote  Hermitian transpose, transpose, matrix trace operator, $N \times N$ identity matrix, all-zero matrix, 
and diagonal matrix, respectively. $\ex{\cdot}$ is the expected value operator.
% and  $(x)^{+}$ is the positive part of $x$, equal to $\max(0,x)$ (elementwise for matrices). 
% $\A \succeq \B$ denotes the Loewner partial order, meaning that $\A - \B$ is positive semidefinite.

\section{Satellite Relaying  Operation}\label{sec:relay}

The satellite serves $K$ terminals at each channel use\footnote{The focus of this study is on the forward link from gateways to user terminals, without precluding the support  to the return link.}. 
All $K$ users get access to the  same frequency spectrum, thus giving rise to both intra-cluster 
and inter-cluster interference. The satellite  has $N$ radiation elements or feeds, with $N \geq K$. 
The $K \times 1$ vector comprising the values received by the $K$ users at a given time instant is written as 
\begin{equation}
\y = \widetilde{\H} \widetilde{\B} \x + \n
\label{eq:initial_model}
\end{equation}
for $K \times 1$ vector $\x$ transmitted values and $K\times 1$ vector $\n$ zero-mean unit variance Additive White Gaussian Noise (AWGN),
 such that 
$\ex {\n \n^H } =\mathbf{I}_K$. The BFN weights are included in matrix  $\widetilde{\B} \in \mathbb{C}^{N \times K}$. 
$\widetilde{\H} \in \mathbb{C}^{K \times N} $ is the overall  user link  channel matrix whose element $[\widetilde{\H}]_{ij}$ 
represents the  gain of the link between the $i$-th user  and the $j$-th satellite  feed. \\

As shown in Figure \ref{fig:system_des}, the number of transmit ground stations is $M$, 
each sending $k$ signal streams  simultaneously (in different frequency slots, for example) to the satellite, 
which makes use of $n$ antenna feeds  to send those symbols to the $k$ users in the $m$th cluster, with $k \leq n \leq N$. 
The groups of $n$ feeds are not necessarily disjoint. The information transmitted from each ground station is written as  $\x_m = \T_m \s_m$, 
with $\T_m \in \mathbb{C}^{k \times k}, m=1,\ldots,M$, a set of distributed precoding matrices, and 
$\s_m \in \mathbb{C}^{k \times 1}, m=1,\ldots,M$ the symbols transmitted by the $m$th gateway. 
The initial model \eqref{eq:initial_model} can be detailed as 
\begin{equation}
\y = \widetilde{\H} \left[ 
\begin{array}{ccc}
\widetilde{\B}_1 & \cdots & \widetilde{\B}_M
\end{array}
\right]
\left[ 
\begin{array}{c}
\T_1 \s_1 \\
\vdots \\
\T_M \s_M
\end{array}
\right]
+ \n
\label{eq:initial_model2}
\end{equation}
where the tall matrices $\widetilde{\B}_m \in \mathbb{C}^{N\times k}$ contain the BFN weights assigned to gateway $m$, and $k\cdot M = K$. 
The  input energy is normalized as   \begin{math}\ex {\s_m \s_m^H } =\mathbf{I}_k \end{math}. 
The goal  of the precoder at each transmitter is mainly the mitigation of  the intra-cluster interference, while trying to reduce the 
 negative impact of its  interference on other clusters. The BFN should exploit the additional degrees of freedom to gain inter-cluster interference and/or noise  resilience, preferably in a robust way against the uncertain location of the users.\\

More specifically, the notation can reflect that only a portion of feeds is involved per cluster. If $n$ denotes the number of feeds serving each cluster, 
the weights with content in the BFN can be collected by the tall submatrices  $\B_m \in \mathbb{C}^{n \times k}, m=1,\ldots,M$,  with  $k \leq n \leq N$.  
Each matrix $\widetilde{\B}_m$ in \eqref{eq:initial_model2} only has $n$ non-zero rows, so we can write 
\begin{equation}
\widetilde{\B}_m = \mathcal{S}_m  \B_m
\end{equation}
 where $\mathcal{S}_m$ comprises $n$ columns of $\I_N$, in particular those with the indices of the feeds used by gateway $m$. 
With $n > k$, there are extra degrees of freedom to fight the inter-cluster interference and gain noise resilience 
without increasing the bandwidth of the user link. If we decompose the received signal and noise vectors in \eqref{eq:initial_model} into their 
respective vectors per cluster, $\y_m$ and $\n_m$, respectively,  
and $\H_{mp}$ is the channel between the $n$ feeds operated by the $p$th gateway and the $m$th cluster, then 
the initial signal model reads as 
\begin{equation}\label{eq:signalmodel}
\begin{pmatrix}
 \mathbf{y}_{1}\\\vdots\\\mathbf{y}_{M}
 \end{pmatrix}= 
\underbrace{
\begin{pmatrix}
 \mathbf{H}_{11} & \hdots & \H_{1M} \\
\vdots  & \ddots & \vdots \\
\H_{M1} &  \hdots & {\mathbf{H}}_{MM}
 \end{pmatrix}}_{\H}
\begin{pmatrix}
 \B_1 &  &  \\
  & \ddots &  \\
   &  & \B_M
 \end{pmatrix}
\begin{pmatrix}
 \T_1 \s_1\\ \vdots \\ \T_M \s_M 
 \end{pmatrix}+
  \begin{pmatrix}
 \mathbf{n}_{1}\\ \vdots \\\mathbf{n}_{M}
 \end{pmatrix}.
\end{equation}
Note that, in this cluster-oriented notation, the channel matrix $\H \in \mathbb{C}^{K \times nM}$ does not coincide with $\widetilde{\H}$ in 
\eqref{eq:initial_model2}. If $\H$ is decomposed as $\left[ \begin{array}{ccc}\H_1 & \cdots & \H_M \end{array} \right]$, and since 
both expressions \eqref{eq:signalmodel} and \eqref{eq:initial_model2} need to be equivalent, we can readily conclude that 
$\H_m = \widetilde{\H} \mathcal{S}_m$. \\

\noindent The vector of samples received by users in cluster $m$ is decomposed as 
\begin{equation}
\y_m = \underbrace{\H_{mm} \B_m \T_m \s_m}_\text{intra-cluster} + \underbrace{\sum_{p \neq m} \H_{mp} \B_p \T_p \s_p}_\text{inter-cluster} + 
\underbrace{\n_m}_\text{noise}.
\end{equation}
End users cannot cooperate, so that we consider a receiver of the form $\hat \s_m = \D_m \y_m$, where the matrix $\D_m$ is $k\times k$ diagonal. The 
particular case in which the same scaling is used across the cluster will be also assumed later, with $\D_m = \frac{1}{\sqrt{t_m}} \I_k$. \\

\noindent As performance metric we  use the aggregated MSE, or Sum MSE (SMSE),  given by 
\begin{equation}\label{eq:SMSEbasic}
\smse = \sum_{m=1}^M \trace\left\{ \E_m \right\},
\end{equation}
with 
\begin{equation}\label{eq:Em}
\E_m \triangleq \ex{(\s_m - \hat \s_m) (\s_m - \hat \s_m)^H}.
\end{equation}
The expectation is  computed with respect to the symbols and the noise for a fixed channel, and reads as 
\begin{multline}\label{eq:SMSE}
\smse = \sum_{m=1}^M \trace \left\{ \I_k - \D_m \H_{mm} \B_m \T_m - \T_m^H \B_m^H \H_{mm}^H \D_m^H  \right. + \\ 
\left. \T_m^H \B_m^H \left( \sum_{p=1}^M \H_{pm}^H \D_p^H \D_p \H_{pm} \right) \B_m \T_m +
 \D_m \D_m^H \right \}, 
\end{multline}
written in such a way that the impact of $\T_m$ and $\B_m$ on the overall error is limited to the $m$th term of the summation. 
This way of dealing together with the interference posed on the same cluster and leaked to other clusters have been explored 
in other works such as \cite{Sayed2007}, where the Signal to Leakage and Noise Ratio (SLNR) was maximized for a single transmitter. Even further, 
\cite{Doufexi2012} showed that the minimization of the MSE and the maximization of SLNR lead to equivalent solutions for equal allocation of power for all users, a single base station and single user terminals. 
It is important to remark that the minimization of SMSE and the maximization of sum capacity are related, 
although they can suffer from lack of fairness issues with less favored users \cite{Vandendorpe2010}. \\

\noindent For convenience, we define 
\begin{equation}\label{eq:AX}
\A_m \triangleq \sum_{p=1}^M \H_{pm}^H \D_p^H \D_p \H_{pm} \in \mathbb C^{n\times n}, \qquad \X_m \triangleq \H_{mm}^H \D_m^H \in \mathbb C^{n \times k}, 
\end{equation}
so that 
\begin{equation}\label{eq:SMSEcomp}
\smse = \sum_{m=1}^M \trace \{ \I_k - \X_m^H \B_m \T_m - \T_m^H \B_m^H \X_m  + \T_m^H \B_m^H \A_m \B_m \T_m + \D_m \D_m^H  \}.
\end{equation}
The SMSE in \eqref{eq:SMSE}, or alternatively \eqref{eq:SMSEcomp},  
is the starting point to explore several solutions for the precoding matrices $\{\T_m\}_{m=1}^M$ and BFN weights $\{\B_m\}_{m=1}^M$, 
each targetting different constraints. 
Under the proposed global MSE framework, all the involved coefficients in the transmission process would be the result of 
minimizing the overall MSE:
\begin{equation}\label{eq:P1}
 \text{(P1)} \qquad \{\T_m,\B_m,\D_m\}_{m=1}^M  =  \arg \min \sum_{m=1}^M \trace\{\E_m\} \qquad
\text{s. to }  \trace\{\B_m \T_m \T_m^H \B_m^H\} \leq P_m, 
\end{equation}
with  $P_m$ the power allocated  to the $m$-th cluster. 
%This power is given by 
%$\sum_{m=1}^M \trace\{\ex{\B_m \T_m \s_m \s_m^H \T_m^H \B_m^H }\} = \sum_{m=1}^M \trace\{\B_m \T_m \T_m^H \B_m^H\}$. 
This is the power limit for each group of $n$ antenna feeds; 
note that we are not considering per-feed-constraints at the satellite, left for future studies. The overall available power on the satellite 
is $P = \sum_{m=1}^M P_m$.\\

\noindent We are interested in addressing the separate optimization of $\{\T_m\}_{m=1}^M$ and $\{\B_m\}_{m=1}^M$, 
since the flexibility and amount of CSI is not necessarily the same on-board the satellite and on-ground. Only for the OGBF case, with all weights operated at the gateways, a joint $\{\B_m \T_m\}_{m=1}^M$ matrix is considered. 

No closed-form seems to be available for variables $\{\D_m,\T_m,\B_m\}_{m=1}^M$ minimizing the SMSE in \eqref{eq:P1} under transmit 
power constraints, so a multistage approach is explored. Initially we will  assume that 
perfect knowledge is available to obtain the optimum weights, with practical constraints being imposed later to come up with a fixed BFN and 
no exchange of signalling information among gateways.

\section{On-Ground Beamforming}\label{sec:OGBF}

For setting  a reference we  start with the most favorable case, for which CSIT  is perfectly known and all coefficients can be correspondingly adjusted.  Joint adaptation of precoding and beamforming coefficients can be applied if their combined operation takes place at the ground stations, by using the OGBF technology detailed in the Introduction. All coefficients can be directly manipulated on the ground, at the price of a higher number of exchanged signals with the satellite, one per feed managed by the corresponding gateway. \\

The grouping of $\B_m$ and $\T_m$ in \eqref{eq:SMSE} leads us to write $\F_m \triangleq \B_m \T_m$, and optimize 
directly with respect to these $\F_m$ matrices. Since no closed-form solution seems to be
available, we optimize cyclically   with respect to $\{\F_m\}_{m=1}^M$ keeping $\{\D_m\}_{m=1}^M$ fixed, 
and then with respect to $\{\D_m\}_{m=1}^M$  keeping $\{\F_m\}_{m=1}^M$ fixed. In this way, convergence in the
cost is guaranteed \cite{Stoica2004}. With fixed $\{\D_m\}_{m=1}^M$, the optimization decouples into $M$ separate problems, for 
$m=1,\ldots,M$: 
\begin{equation}\label{eq:OGBF}
 \text{(P2)} \quad \F_m  =  \arg \min \trace \{ \I_k  - \X_m^H \F_m - \F_m^H \X_m  + \F_m^H \A_m \F_m + \D_m \D_m^H \}  \quad
\text{s. to }  \trace\{\F_m \F_m^H\} \leq P_m.
\end{equation}
The minimization of \eqref{eq:OGBF} reduces to a Least Squares problem with a quadratic inequality constraint. 
The solution is found as follows,  for $m=1,\ldots,M$ and $M>1$:
\begin{itemize}
\item First, check whether the unconstrained solution $\F_m = \A_m^{-1} \X_m$ is  feasible; if so, stop. 
\item  Otherwise, the constraint is satisfied  with equality; the solution is $\F_m = (\A_m + \nu_m \I_n)^{-1} \X_m$, 
where the Lagrange multiplier $\nu_m$  has to be numerically computed to meet the power constraint $\trace\{\F_m \F_m^H\} = P_m$, see the Appendix. 
\end{itemize}
Now we fix the matrices $\{\F_m\}_{m=1}^M$ and need to find $\{\D_m\}_{m=1}^M$. We rewrite the SMSE \eqref{eq:SMSE} as 
\begin{multline}\label{eq:SMSErew}
\smse = \sum_{m=1}^M \trace \left\{ \I_k - \D_m \H_{mm} \F_m - \F_m^H  \H_{mm}^H \D_m^H  \right. + \\ 
\left. \D_m \left( \sum_{p=1}^M \H_{mp} \F_p \F_p^H \H_{mp}^H \right) \D_m^H +
 \D_m \D_m^H \right \}
\end{multline}
and define
\begin{equation}
\C_m \triangleq \I_k + \sum_{p=1}^M \H_{mp} \F_p \F_p^H \H_{mp}^H \in \mathbb C^{k\times k}, \qquad \G_m \triangleq \F_m^H \H_{mm}^H \in \mathbb C^{k \times k},
\end{equation}
so that we obtain the following compact expression for the SMSE:
\begin{equation}\label{eq:SMSE_Dm}
\smse = \sum_{m=1}^M\trace \{ \I_k - \D_m \G_m^H - \G_m \D_m^H + \D_m \C_m \D_m^H\}.
\end{equation}
The minimization of \eqref{eq:SMSE_Dm} subject to $\D_m$ being diagonal is straightforward: if we let 
$\D_m = \text{diag} \{\begin{array}{ccc} d_1^{(m)} & \cdots & d_k^{(m)} \end{array} \}$, then 
\begin{equation}
d_j^{(m)} = \frac{[\G_m]_{jj}}{[\C_m]_{jj}}, \qquad j=1,\ldots,k, \quad m=1,\ldots,M. 
\end{equation}

The {\bf single gateway case ($M=1$)} offers the best possible performance since all streams can have access to all feeds, achieving  a better attenuation of the  co-channel interference. In the above computations it must be noted that $\A_m$ is no longer full-rank; from \eqref{eq:AX}, we have that the 
rank of $\A_m$ is not higher than $K$. The unconstrained solution which must be tested first for feasibility is $\F = \A^\dagger \X$, with $\A^\dagger$ the 
pseudo-inverse of $\A$, and subscript $m$ dropped. Even further, if 
the scaling parameter is the same for all terminals, with $\D = (1/\sqrt{t}) \I_K$, then the previous mathematical derivations can be simplified. Thus, 
$\F$ in \eqref{eq:OGBF} is simply given by 
\begin{equation}
\F = \sqrt{t}(\H^H \H + \gamma \I_K)\H^H
\end{equation}
and $\gamma = K/P$. This result is already reported in \cite{DevillersNeiraMosquera}, and can be proved by using the eigen-value decomposition of the channel Gramian $\H^H \H = \U \S \U^H$ and similar steps to those exposed in the next section.

\section{On-Board Beamforming}\label{sec:OBBF}

On-board beamforming needs only the exchange of one stream per user, not per feed, and it is the most common in practice, 
with different degrees of flexibility. Fully adaptive OBBF weights turn  out to be 
highly  challenging from the implementation point of view and, as a general rule, the adaptation time scale of BFN weights is more constrained than that of ground precoding weights \cite{Tronc2014}. 
This is why we split the adaptation of both sets of coefficients in this section, 
in an effort to leverage their separate roles and eventually design 
a fixed BFN or with a limited degree of programmability. \\

The complexity of (P1) is such that no closed-form expressions can be jointly obtained. 
For convenience, we  assume that the beamforming matrices are semi-unitary, with orthonormal columns:
\begin{equation}
\B_m^H \B_m = \I_k, m=1,\ldots,M.
\end{equation}
Any rank-$k$ $\B_m \in \mathbb{C}^{n\times k}$ can be non-uniquely factorized as $\B_m = \B_m^0 \R$, with invertible $\R \in \mathbb{C}^{k\times k}$ 
and $\B_m^0 \in \mathbb{C}^{n\times k}$ with orthonormal columns. As beamforming weights we choose $\B_m^0$, and $\R$ can be embedded in the precoder 
$\T$ without affecting the minimum MSE. With this, the power constraint in (P1) can be written as $\trace\{\T_m \T_m^H \} \leq P_m$. \\
If we fix the beamforming weights, then the solution of the previous section applies, by using $\B_m^H \A_m \B_m$ and $\B_m^H \X_m$ in lieu of 
$\A_m$ and $\X_m$, respectively:
\begin{equation}\label{eq:generalTm}
\T_m = \left(\B_m^H \A_m \B_m + \nu_m \I_k \right)^{-1} \B_m^H \H_{mm}^H \D_m^H.
\end{equation}
Some of the Lagrange multipliers $\{\nu_m\}_{m=1}^M$ could be zero if the power constraint of the $m$-th cluster is not active. 
Note that the different scaling matrices  $\{\D_m\}_{m=1}^M$ are also embedded in $\bm \A_m$ as per \eqref{eq:AX}. Again, 
an iterative process could serve to iterate till convergence $\{\T_m\}_{m=1}^M$ and  $\{\D_m\}_{m=1}^M$. In the particular case in which 
$\D_m = \frac{1}{\sqrt{t_m}} \I_k$, i.e., the same scalar applies for all users belonging to the same cluster, then  $\T_m$ is written as 
\begin{equation}\label{eq:Tminter}
\T_m = \sqrt{t_m} \left( \B_m^H (t_m\A_m) \B_m + \nu_m' \I_k  \right)^{-1} \B_m^H \H_{mm}^H
\end{equation}
where $\nu_m' = t_m \nu_m$. $\A_m$, the sum of the  channel Gramians  from those feeds managed by the $m$th gateway to users in all clusters, can 
be expressed, from \eqref{eq:AX}, as 
\begin{equation}\label{eq:fullAm}
\A_m =  (1/t_m)(\H_{mm}^H \H_{mm} + \bm \Sigma_m)
\end{equation}
with 
\begin{equation}
\bm \Sigma_m \triangleq \sum_{\substack{p=1 \\p \neq m}}^M \frac{t_m}{t_p} \H_{pm}^H \H_{pm}.
\label{eq:Sigmam}
\end{equation}
% For fixed  beamforming weights, the precoding matrices $\T_m,m=1,\ldots,M$, operated at each gateway as in 
% \eqref{eq:signalmodel},  are obtained as 
% \begin{equation}\label{eq:P3}
% \text{(P3)} \qquad \{\T_m\}_{m=1}^M  =  \arg \min \sum_{m=1}^M \trace\{\E_m\} \qquad
% \text{s. to }  \trace\{\T_m \T_m^H\} \leq P_m. 
% \end{equation}
% By defining the Lagrangian 
% \begin{equation}
% {\cal L}\left( \{\T_m\},\{\bm \nu_m\} \right) = \sum_{m=1}^M \trace\{\E_m\} + 
% \sum_{m=1}^M \nu_m \left( \trace\{\T_m \T_m^H\}- P_m \right) 
% \end{equation}
% and equating to zero $\partial {\cal L} \left( \{\T_m\},\{\bm \nu_m\}\right)/\partial \T_m = 0$, 
% we get, for $m=1,\ldots,M$, 
% since $\T_m$ is only involved in the $m$-th error term $\trace\{\E_m\}$. The Lagrange multipliers $\{\nu_m\}_{m=1}^M$ can be considered as  
% {\bf regularization weights} accounting for the relative contribution of noise and interference. 
%   and needs to be obtained such that the contribution of the $m$-th gateway to the SMSE is minimized.
The first term in $\A_m$ corresponds to the intra-cluster channel, whereas the second collects the leakage channel to all other clusters. 
For practical reasons, the acquisition of the  inter-cluster channels to make  $\bm \Sigma_m$ available to gateway $m$ is difficult to 
guarantee in practice. Even in the case that $\H_{pm}$ were known, the scaling parameters  $\{t_m\}$ present  in  $\bm \Sigma_m$ would 
need coordination for their computation; 
a sequential  process, for instance, would obtain  $\{\nu_m'\}$ for an initial set of $\{t_m\}$, set which 
would be recomputed for the obtained values of $\{\nu_m'\}$, and so on. 
Message passing algorithms such as in \cite{Ting2014} could be devised for this process, although  are left outside the scope of this paper, 
which hinges on the autonomous operation of the gateways. To the end of simplifying the implementation, we will consider that 
$\B_m^H \bm \Sigma_m \B_m$ is approximated by $c_m \cdot \I_K$, with $c_m$ a constant which can be absorbed 
by the regularization factor. 
Thus, by reducing $\B_m^H (t_m \A_m)  \B_m$ to $\B_m^H \H_{mm}^H \H_{mm} \B_m + c_m \I_k$, 
 the precoder of the $m$th gateway is 
\begin{equation}\label{eq:Tm}
\T_m = \sqrt{t_m} \left( \B_m^H \H_{mm}^H  \H_{mm} \B_m + \gamma_m \I_k  \right)^{-1} \B_m^H \H_{mm}^H,
\end{equation}
with $\gamma_m = \nu_m'+c_m$. The {\bf regularization factors} $\gamma_m$  need to be obtained so that the contribution of the $m$-th 
gateway to the 
SMSE is minimized, as addressed in the following sections. If $\gamma_m = k/P_m$, this is the intra-cluster MMSE precoder \cite{DevillersNeiraMosquera}. 
Note that the structure 
of the precoder is the same as that obtained from considering $\bm \Sigma_m = \0$. 
A variant which operates with the full $\A_m$ matrix in 
\eqref{eq:fullAm} is presented in \cite{Mosquera17submitted}, and it is left out of the scope of this work. \\

As stated eariler, we need to point out that the separated optimization of the ground precoders $\{\T_m\}_{m=1}^M$ and the BFN $\{\B_m\}_{m=1}^M$ 
has as ultimate goal to fix the BFN and let the precoders adapt to the channel variations. 
This is why we do not pursue the full optimization of $\B_m$ in (P1)  for fixed 
$\T_m$ and enter into a sequential minimization process as that in Section \ref{sec:OGBF}, 
but instead try to decouple the derivation of $\B_m$ from $\T_m$. Next we illustrate how to obtain the scaling $t_m$ and 
the regularization factor $\gamma_m$ in \eqref{eq:Tm} under different restrictions on $\B_m$.

\subsection{Pre-fixed BFN} 

If the $\{\B_m\}$ weights are already in-place and cannot be altered, then the parameters to optimize are 
$\{t_m,\gamma_m\}$ in \eqref{eq:Tm}.  The problem can be posed  as 
\begin{equation}\label{eq:fixedBFN}
\text{(P3)} \qquad \{t_m,\gamma_m\}_{m=1}^M  =  \arg \min  \trace\{\E_m\} \qquad
\text{s. to }  \trace\{\T_m \T_m^H\} \leq P_m. 
\end{equation}
This error is written as 
\begin{multline}\label{eq:Em}
\trace\{\E_m\} = \tr\{\I_k\} - 2 \tr\{ (\B_m^H \H_{mm}^H \H_{mm} \B_m + \gamma_m \I_k )^{-1} \B_m^H \H_{mm}^H \H_{mm} \B_m\} \\ 
+ \tr\{(\B_m^H \H_{mm}^H \H_{mm} \B_m + \gamma_m \I_k)^{-1} \B_m^H ( \H_{mm}^H \H_{mm} + \bm \Sigma_m) \B_m 
(\B_m^H \H_{mm}^H \H_{mm} \B_m + \gamma_m \I_k)^{-1} \cdot \\
 \B_m^H \H_{mm}^H \H_{mm} \B_m \} + \tr\left\{ \frac{1}{t_m} \I_k\right\}.
\end{multline}
Note that all $M$ minimization problems in (P3) are coupled through the inter-cluster term $\bm \Sigma_m$. 
They can be decoupled if we assume that all $\{t_m\}$ are similar, so that  message exchange among the gateways 
can be avoided during the optimization phase:
\begin{equation}\label{eq:approx}
\bm \Sigma_m \approx \sum_{\substack{p=1 \\p \neq m}}^M \H_{pm}^H \H_{pm}.
\end{equation}
 This looks like a reasonable assumption for large numbers of users, as the results in Section 
\ref{sec:simulations} will show. As a consequence, the power restrictions in \eqref{eq:fixedBFN} become active, with $t_m$ taking the 
highest  possible value.\\ 

With the $M$ problems decoupled, the regularization factor $\gamma_m$ at each gateway precoder \eqref{eq:Tm} can be designed so that the contribution $\trace\{\E_m\}$ to the 
SMSE is minimized. If this can be effectively applied, then  the resulting precoder will be inter-cluster aware. The optimum value is given by the 
following lemma, which is proved in the Appendix.  
\begin{lemma}
If we write the eigen-decomposition 
\begin{equation}
\B_m^H \H_{mm}^H \H_{mm }\B_m = \U_m \S_m \U_m^H
\end{equation}
with $\S_m = \text{diag}\{\begin{array}{ccc}\lambda_1^{(m)} & \cdots & \lambda_k^{(m)} \end{array}\}$, 
  then the regularization factor $\gamma_m$ minimizing $\trace\{\E_m\}$ 
is the solution of the following equation:
\begin{equation}\label{eq:gamma-fixed}
\sum_{i=1}^k \frac{\lambda_i^{(m)}}{(\lambda_i^{(m)}+\gamma_m)^3}\left(\gamma_m - \sigma_{ii}^{(m)} - \frac{k}{P_m}\right) = 0
\end{equation}
with $\sigma_{ii}^{(m)}$  the $i$th diagonal entry of $\U_m^H \B_m^H \bm \Sigma_m \B_m \U_m$.  
The corresponding  scaling parameter of the precoder and receiver is given by 
\begin{equation}\label{eq:tm-fixed}
t_m = P_m / \sum_{i=1}^k \frac{ \lambda_i^{(m)}}{(\lambda_i^{(m)} + \gamma_m)^2}. 
\end{equation}
\end{lemma}

If all $\sigma_{ii}^{(m)}$ are equal to zero -no inter-cluster interference- then the solution to \eqref{eq:gamma-fixed} 
is trivially seen to be $\gamma_m = k/P_m$, similarly to \cite{Hochwald2005a}. More generally, 
it can be readily seen to  lie in the interval  $[k/P_m, k/P_m + \max(\sigma_{ii}^{(m)})]$. 
However, its derivation relies  on the knowledge of $\bm \Sigma_m$, which participates in \eqref{eq:gamma-fixed} through $\sigma_{ii}^{(m)}$. 
Even though we use  the approximation in \eqref{eq:approx}, the lack of coordination  among gateways prevents 
the acquisition of the channel response from 
feeds serving cluster $m$ to terminals in all other clusters; we propose instead to make use of the expected leakage channel Gramians, thus avoiding 
their instantaneous acquisition. With this,  we approximate \eqref{eq:Sigmam} as 
\begin{equation}\label{eq:approx1}
\bm \Sigma_m \approx \hat{\bm \Sigma}_m = \sum_{\substack{p=1 \\p \neq m}}^M \ex{\H_{pm}^H \H_{pm}}.
\end{equation}
%If the expectation of the inter-cluster channel Gramians can be computed off-line, it will contribute to mitigate the 
%the summation across clusters is expected not to differ too much. 
%If exchange of information among gateways has to be avoided, and it is not possible to acquire the channel response from , a more pragmatic regularization% factor can be obtained from the following ad-hoc design:
%\begin{equation}\label{eq:pragmatic}
%\gamma_m = k/P_m + \trace\{\B_m^H \ex{\bm \Sigma_m} \B_m\}/k,  
%\end{equation}
%provided the expectation can be estimated off-line. The summation across clusters when building $\bm \Sigma_m$ is expected to have a beneficial 
%effect, in the sense of averaging out the errors coming from the expectation approximation. 
In addition to the numerical solution of \eqref{eq:gamma-fixed}, we will also test in the simulations the following approximation: 
\begin{equation}\label{eq:approx-gamma}
\gamma_m = k/P_m + \trace\{\B_m^H \hat{\bm \Sigma}_m \B_m \}/k. 
\end{equation}
The first term $k/P_m$ is the regularization factor for intra-cluster precoders; the second term comes from approximating 
$\B_m^H \bm \Sigma_m \B_m$ by  $c_m \I_k$ in \eqref{eq:Tm}, in such a way that the trace of both matrices is the same
(for identical  $\{t_m\}$ values  in $\bm \Sigma_m$). The precoder computed in this way  is still inter-cluster aware, yielding an edge with respect to intra-cluster 
precoders. We will show the validity of this approach in the simulations.

\subsection{Adaptive BFN}\label{sec:adaptiveBFN}

If the BFN weights can be optimized, at first sight we should  choose $\B_m$ under the SMSE criterion to minimize  the error term $\trace\{\E_m\}$ in (P1) as
\begin{eqnarray}\nonumber
\text{(P4)} \quad  \{\B_m\}_{m=1}^M  & =    \arg \min \trace\{\I_k - \frac{1}{\sqrt{t_m}}\left(\H_{mm} \B_m \T_m  - 
\T_m^H \B_m^H \H_{mm}^H \right) + 
\T_m^H \B_m^H \A_m \B_m \T_m  + \frac{1}{t_m} \I_k  \} \\
& \text{s. to} \quad \B_m^H \B_m = \I_k.
\label{eq:minBm}
\end{eqnarray}
Desirably,  we would like to decouple the derivation of  BFN $\B_m$ and the precoder $\T_m$ as much as possible to simplify the practical implementation, so that different degrees of flexibility can be accommodated. 
 No closed-form seems 
to be feasible for $\B_m$ minimizing all three components together. For the zero-forcing version of the precoder $\T_m$, that is, with $\gamma_m=0$ in \eqref{eq:Tm}, the inter-cluster contribution in (P4) becomes independent of $\B_m$, and only remain the inter-cluster interference and additive noise components.  
If $\B_m$ is designed to minimize the effect of the latter, then $t_m$ in (P4) needs to be maximized.  This problem is written now as 
% \begin{eqnarray}\label{eq:costnoise}
% \text{(P5)}  \quad  \{\B_m,t_m\}_{m=1}^M  & =  \arg \min \trace \{\frac{1}{t_m}\I_k\} =  \arg \min \trace\{(\B_m^H \H_{mm}^H \H_{mm} \B_m)^{-1}\} \\
% & \text{s. to} \quad \B_m^H \B_m = \I_k.
% \nonumber
% \end{eqnarray}
\begin{eqnarray}\label{eq:costnoise}
\text{(P5)}  \quad  \{\B_m,t_m\}_{m=1}^M  & = & \arg \min \trace \left\{\frac{1}{t_m}\I_k\right\}\\
& \text{s. to} & \quad \left\{ \begin{array}{ll} \B_m^H \B_m = \I_k, \\ 
\trace\{\T_m \T_m^H\} \leq P_m \end{array} \right.
\nonumber
\end{eqnarray}
with $\T_m = \sqrt{t_m}(\B_m^H \H_{mm}^H \H_{mm} \B_m)^{-1}\B_m^H \H_{mm}^H$. It can be readily seen that, at the optimum point, 
the power constraint must hold with equality, and 
\begin{equation}
t_m =  P_m/\trace\{\T_m \T_m^H\} = P_m/\trace\{(\B_m^H \H_{mm}^H \H_{mm} \B_m)^{-1}\}.
\end{equation}
This design scheme effectively decouples the derivation of BFN and precoder, and exploits the degrees of freedom available at the satellite to increase the resilience against the noise. 
As a remark, the inter-cluster leakage will be  again addressed by the proper design  of the regularization factor as shown later. 
As noted in the previous paragraph for fixed $\B_m$, 
the regularization factor $\gamma_m$ in \eqref{eq:Tm} has a non-trivial dependence on the channel and the BFN.\\

Let us write  the eigenvalue decomposition  
\begin{equation}\label{eq:decomposeHH}
\H_{mm}^H\H_{mm} = \U_{H,m} \left( \begin{array}{cc} \S_{H,m} & \0 \\ \0 & \0 
\end{array} \right) \U_{H,m}^H,
\end{equation}
with $\S_{H,m} = \text{diag} \{\begin{array}{ccc} \lambda_1^{(m)} & \cdots & \lambda_k^{(m)} \end{array} \}$, and $\lambda_i^{(m)}$ denoting the $k$ non-zero 
eigenvalues 
of $\H_{mm}^H\H_{mm}$ in decreasing order. With this, we have
\begin{equation}
\trace\{(\B_m^H \H_{mm}^H \H_{mm} \B_m)^{-1}\} = \sum_{i=1}^k \frac{1}{\lambda_i(\B_m^H \U_{H,m} 
{\tiny \left( \begin{array}{cc} \S_{H,m} & \0 \\ \0 & \0 \end{array} \right)}
 \U_{H,m}^H \B_m)}
\end{equation}
with $\lambda_i(\Z),i=1,\ldots,k$,  denoting the eigenvalues of $\Z$ in decreasing order. 
$\U_{H,m}^H \B_m$ has orthonormal columns, since $\B_m^H \B_m = \I_k$, so  we can apply 
Poincar\'e separation theorem \cite{Horn13}, which bounds the eigenvalues of 
$\B_m^H \U_{H,m} {\tiny \left( \begin{array}{cc} \S_{H,m} & \0 \\ \0 & \0 \end{array} \right)} \U_{H,m}^H \B_m$ in terms of those of 
${\tiny \left( \begin{array}{cc} \S_{H,m} & \0 \\ \0 & \0 \end{array} \right)}$ in the following way:
\begin{equation}
\lambda_i^{(m)} \geq \lambda_i(\B_m^H \U_{H,m} {\tiny \left( \begin{array}{cc} \S_{H,m} & \0 \\ \0 & \0 \end{array} \right)} \U_{H,m}^H \B_m) \geq \lambda_{n-k+i}^{(m)}
\end{equation}
so that 
\begin{equation}
\sum_{i=1}^k \frac{1}{\lambda_i(\B_m^H \U_{H,m} {\tiny \left( \begin{array}{cc} \S_{H,m} & \0 \\ \0 & \0 \end{array} \right)} \U_{H,m}^H \B_m)} \geq \sum_{i=1}^k \frac{1}{\lambda_i^{(m)}}.
\end{equation}
The lower bound is achieved for 
$\B_m^H \U_{H,m} = [ \begin{array}{cc}\I_k & \0\end{array}]$, so the optimal solution  $\B_m$ must be equal to the first $k$ columns of $\U_{H,m}$. 
With this solution the $n-k$ degrees of freedom provided by $\B_m$ are exploited to reduce the impact of noise enhancement 
due to the intra-cluster cancellation. 

The design of the BFN is such that $\B_m^H \H_{mm}^H \H_{mm} \B_m = \S_{H,m}$, with $\S_{H,m}$ the  diagonal matrix containing the 
$k$ non-zero eigenvalues of the channel Gramian 
$\H_{mm}^H \H_{mm}$ in \eqref{eq:decomposeHH}, so the precoder matrix reads as 
\begin{equation}\label{eq:Tm_diag}
\T_m = \sqrt{t_m} \left( \S_{H,m} + \gamma_m \I_k  \right)^{-1} \B_m^H \H_{mm}^H.
\end{equation}
As in the case for fixed $\B_m$,  the regularization factor $\gamma_m$ at each gateway precoder \eqref{eq:Tm_diag} 
can be designed so that its contribution $\trace\{\E_m\}$ to the 
SMSE is minimized. Again, the solution for $t_m$ and $\gamma_m$ is that for the fixed case in 
\eqref{eq:gamma-fixed} and \eqref{eq:tm-fixed}. 
With respect to the solution for an isolated cluster, $k/P_m$,  the optimized 
regularization factor is higher  to account for the inter-cluster leakage. 
We will see in the simulations that the properly chosen  increment of  the regularization factor is critical for the performance of the system. 
Again, we propose to resort to 
the approximation \eqref{eq:approx1} to avoid the communication among gateways. 
% We propose the following ad-hoc expression for $\gamma_m$ as an extension of that in \eqref{eq:pragmatic}, which applied for fixed $\B_m$:
% \begin{equation}\label{eq:pragmaticE}
% \gamma_m = k/P_m + \trace\{\ex{\B_m^H \bm \Sigma_m \B_m}\}/k. 
% \end{equation}

\subsection{Null steering}
Some or all the degrees of freedom of $\B_m$ can be used to cancel the interference posed by the $m$th gateway on some given off-cluster users. 
Inter-cluster cancellation was also addressed  in \cite{Devillers2011}, in this case from the ground in the absence of on-board BFN. In 
our setting the ground precoders can follow the design in the previous sections, and the on-board BFN can create nulls in some specific locations. 
The off-cluster locations to preserve free of interference  could be fixed or time-varying provided that some mechanism exists to track the corresponding channels. \\
 If $\bar k$ denotes the number of users which must be protected, 
then we have that $\bar k \leq n - k$. The rows of the matrix $\H$ containing the channel from the feeds allocated to 
the  $m$th gateway   to those  selected $\bar k$ users 
  are collected under  $\bar \H_{mm} \in {\mathbb C}^{\bar k \times n}$ matrix, assumed to be full rank, 
that is, of rank $\bar k$. 
The elimination of the inter-cluster interference to some users will come at the price of increased overall MSE, due to the reduction of 
available  degrees of freedom to solve (P5), which now reads as 
\begin{eqnarray}
\text{(P6)}  \quad  \{\B_m\}_{m=1}^M  & =  &  \arg \min \trace\{(\B_m^H \H_{mm}^H \H_{mm} \B_m)^{-1}\} \\
& \text{s. to} & \quad \left\{ \begin{array}{ll} \bar \H_{mm} \B_m = \0, \\ 
\B_m^H \B_m = \I_k. \end{array} \right.
\nonumber
\end{eqnarray}
Let the singular value decomposition  of $\bar \H_{mm}$ be expressed as 
\begin{equation}
\bar \H_{mm} = \bar \U_m \bar \S_m \bar \V_m^H,
\end{equation}
with $\bar \U_m \in \mathbb{C}^{\bar k \times \bar k}$,  $\bar \V_m \in \mathbb{C}^{n \times n}$, and 
$
\bar \S_m = \left( \begin{array}{ccc} \ddots & \vline & \0  \end{array} \right) \in \mathbb{C}^{\bar k \times n}.
$
The last $n - \bar k$ columns of $\bar \V_m$ generate the null space of $\bar \H_{mm}$; let us form the 
matrix $\bar \V_m^0 \in {\mathbb C}^{n \times (n-\bar k)}$ with those columns, so the cancellation can be achieved by making use of the null-space projection \cite{Klein2009}, 
\cite{Haardt2004}, building $\B_m$ as 
\begin{equation}
 \B_m  =  \bar \V_m^0  \B_m^0.
\end{equation}
%\begin{equation}
% \B_m  \Bigr|_{n \times k} =  \bar \V_m^0  \B_m^0 
%\end{equation}
Note the reduction in the degrees of freedom, since the number of rows of $\B_m^0 \in \mathbb C^{(n-\bar k) \times k}$ 
gets reduced from $n$ down to $n - \bar k$. If we define 
$\Q_m \triangleq \H_{mm} \bar \V_m^0 \in \mathbb C^{k \times (n - \bar k)}$, then the optimization (P6) is rephrased as 
\begin{eqnarray}
\text{(P7)}  \quad  \{\B_m^0\}_{m=1}^M  & =    \arg \min \trace\{((\B_m^0)^H \Q_m^H \Q_m \B_m^0)^{-1}\} \\
& \text{s. to} \quad (\B_m^0)^H \B_m^0 = \I_k,
\nonumber
\end{eqnarray}
and the derivation of $\B_m^0$ follows, \textit{mutatis mutandis}, the sequence in  problem (P5), by working with $\Q_m$ and $n-\bar k$ 
instead of $\H_{mm}$ and  $n$, respectively.

\subsection{Coarse BFN}\label{sec:robust}

An accurate tracking of the channels to synthesize any of the on-board beamforming solutions exposed above is not easy to implement, especially 
when it is required for a permanent adaptation of the BFN weights on-board the satellite. 
The changes in the channel matrix $\H$ are due, to a large 
extent, to the random relative location of the users with respect to the satellite.  We fix the weights of the satellite BFN
 so that the  
precoders at the gateways undertake all the effort to adapt to the varying CSI, at least partially. 
The regularization factor of the precoders, as shown earlier,  can be also judiciously chosen  to avoid the exchange of information among clusters and simplify the implementation. Based on two premises, (i) the need to fix the BFN weights, and (ii) the absence of exchange of structured information among 
clusters, we apply the results of previous sections by using the expected behavior of the channels when needed. 
The following logic sequence describes the proposed design:
\begin{enumerate}
\item Obtain the fixed BFN as 
\begin{eqnarray}
\text{(P8)}  \quad  \{\B_m\}_{m=1}^M  & =  &  \arg \min \trace\{(\B_m^H \ex { \H_{mm}^H \H_{mm}  } \B_m)^{-1}\} \\
& \text{s. to} &  \quad  \B_m^H \B_m = \I_k. 
\nonumber 
\end{eqnarray}
The solution follows the steps in Section \ref{sec:adaptiveBFN} for the adaptive case, 
with $\ex { \H_{mm}^H \H_{mm}}$ playing the role of $\H_{mm}^H \H_{mm}$. 
\item Adapt the gateway precoders to the channel changes as 
\begin{equation}
\T_m = \sqrt{t_m} \left( \B_m^H \H_{mm}^H  \H_{mm} \B_m + \gamma_m \I_k  \right)^{-1} \B_m^H \H_{mm}^H.
\end{equation}
\item The regularization factor $\gamma_m$ is computed as the solution of 
\begin{equation}
\sum_{i=1}^k \frac{\lambda_i^{(m)}}{(\lambda_i^{(m)}+\gamma_m)^3}\left(\gamma_m - \sigma_{ii}^{(m)} - \frac{k}{P_m}\right) = 0
\end{equation}
with $\sigma_{ii}^{(m)}$  the $i$th diagonal entry of $\U_m^H \B_m^H \hat{\bm \Sigma}_m \B_m \U_m $, and $\hat{\bm \Sigma}_m$ defined in \eqref{eq:approx1}.  
$\{\lambda_i^{(m)}\}$ are the $k$ non-null eigenvalues of $\B_m^H  \H_{mm}^H \H_{mm}   \B_m$.  
The scaling parameter $t_m$ is  such that $\trace\{\T_m \T_m^H\} = P_m$. 
\end{enumerate}

% \item The regularization factor $\gamma_m$ is computed as   
% \begin{equation}
% \gamma_m = k/P_m + \trace\{\B_m^H \sum_{\substack{p=1 \\p \neq m}}^M \ex{\H_{pm}^H \H_{pm}} \B_m\}/k,  
% \end{equation}

% \item For the derivation of the precoder regularization factor $\gamma_m$, \eqref{eq:gamma-fixed} was used, 
%  with $\lambda_i^{(m)}$ the $i$th eigenvalue of $\B_m^H \ex{\H_{mm}^H \H_{mm}} \B_m$, and  
% $\sigma_{ii}$ in \eqref{eq:gamma-fixed} the $i$th diagonal entry of 
% $\B_m^H \ex { \bm \Sigma_m } \B_m$, with $\ex { \bm \Sigma_m }  \approx \sum_{\substack{p=1 \\p \neq m}}^M \ex {\H_{pm}^H \H_{pm} } $. 
% The scaling factor $t_m$ will be again such that $\trace\{\T_m \T_m^H\} = P_m$. 

\noindent All in all, Tables \ref{tab:expressions} and \ref{tab:scalars}  compile the expressions of the proposed transmit precoders and satellite beamforming weights for the 
OBBF case. The coarse OBBF solution, computed as detailed in the previous steps,  is such that the satellite BFN weights are fixed, and the ground transmitters adapt to cope with the intra-cluster interference among their respective users, based on MMSE precoders with a regularization factor tuned to reduce the leakage onto other clusters. 

\begin{table}[]
\centering
\caption{BFN and distributed precoders}
\label{tab:expressions}
\begin{tabular}{|l|l|lll}
\cline{1-2}
\multirow{2}{*}{ {\bf OBBF-adaptive}} & $\B_m$ built as the first $k$  left singular vectors of $\H_{mm}^H \H_{mm}$
   &  &  &  \\ \cline{2-2}
                  &  $\T_m = \sqrt{t_m} \left( \B_m^H \H_{mm}^H \H_{mm} \B_m + \gamma_m \I_k \right)^{-1} \B_m^H \H_{mm}^H$  & & & 
\\ \cline{1-2}
\multirow{2}{*}{ {\bf OBBF-nulling}} &  $\B_m = \bar{\V}_m^0 \B_m^0$,  with $\bar{\V}_m^0$ the null space of $\bar \H_{mm}$, & & &  \\ 
& and $\B_m^0$ built as the first $k$ left singular vectors of $\bar \V_m^{0,H} \H_{mm}^H \H_{mm} \bar \V_m^0$ 
  &  &  &  \\ \cline{2-2}
                  &  $\T_m = \sqrt{t_m} \left( \B_m^H \H_{mm}^H \H_{mm} \B_m + \gamma_m \I_k \right)^{-1} \B_m^H \H_{mm}^H$ 
& & & 
 \\ \cline{1-2}
\multirow{2}{*}{{\bf OBBF-coarse} } & $\B_m$ built as the first $k$  left singular vectors of $\ex { \H_{mm}^H \H_{mm} }$
   &  &  &  \\ \cline{2-2}
                  & $\T_m = \sqrt{t_m} \left( \B_m^H \H_{mm}^H \H_{mm} \B_m + \gamma_m \I_k \right)^{-1} \B_m^H \H_{mm}^H$ & & & 
\\ \cline{1-2}
\end{tabular}
\end{table}

\begin{table}[]
\centering
\caption{Transmitter regularization factor and receiver gain}
\label{tab:scalars}
\begin{tabular}{|l|l|}
\hline
\multicolumn{2}{|c|}{$\gamma_m$ is the solution of  $\sum_{i=1}^k \frac{\lambda_i^{(m)}}{(\lambda_i^{(m)}+\gamma_m)^3}\left(\gamma_m - 
\sigma_{ii}^{(m)} - \frac{k}{P_m} \right) = 0$} \\ \hline
\multicolumn{2}{|c|}{$\sigma_{ii}^{(m)}$ is  the $i$th diagonal entry of $\U_m^H \B_m^H \hat{\bm \Sigma}_m \B_m \U_m $} \\ \hline
{\bf OBBF-adaptive}          & $\lambda_i^{(m)}$ is the $i$th  eigenvalue of  $\B_m^H\H_{mm}^H \H_{mm}\B_m$           \\ \hline
{\bf OBBF-nulling}         & $\lambda_i^{(m)}$ is the $i$th  eigenvalue of  $\B_m^H \Q_m^H \Q_m \B_m$          \\ \hline
{\bf OBBF-coarse}          & $\lambda_i^{(m)}$ is  the $i$th non-null eigenvalue of  $\B_m^H  \H_{mm}^H \H_{mm}   \B_m$           \\ \hline
\multicolumn{2}{|c|}{$t_m = P_m / \sum_{i=1}^k \frac{\lambda_i^{(m)}}{(\lambda_i^{(m)} + \gamma_m)^2} $} \\ \hline
\end{tabular}
\end{table}

\section{Numerical Results}\label{sec:simulations}

\begin{figure}[!h]
\centering
\includegraphics[width=0.65\textwidth]{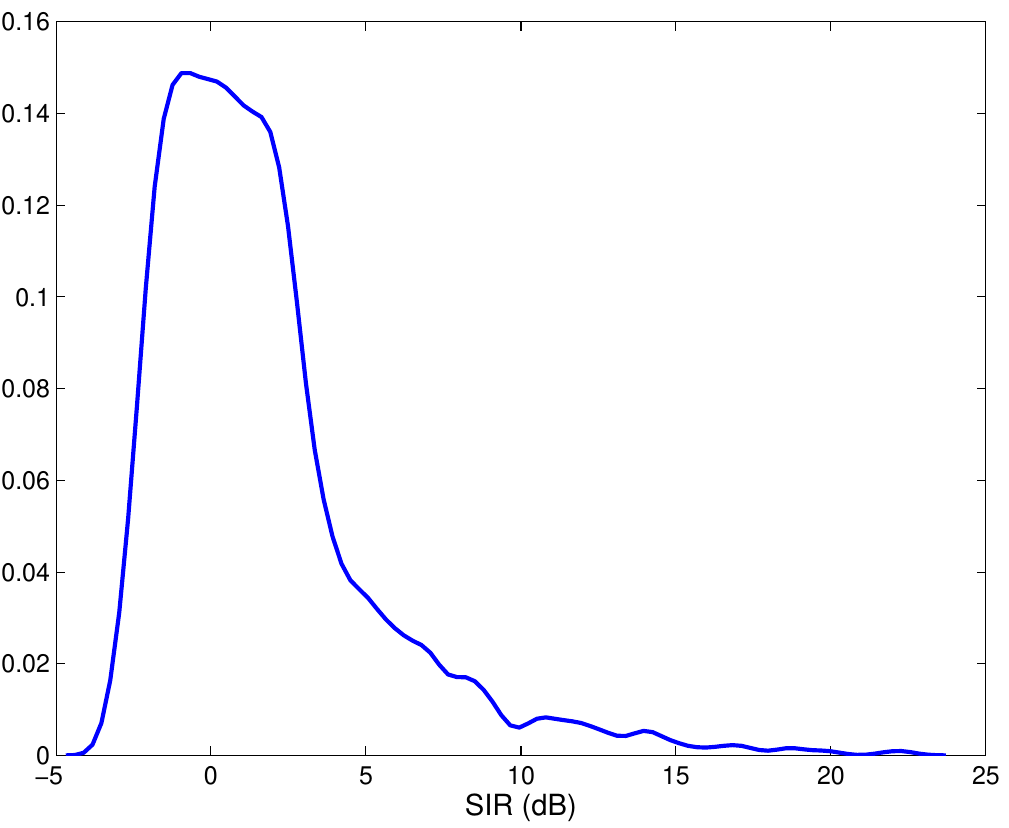}
\caption{SIR histogram without precoding, pre-fixed BFN.}
\label{fig:SIRhistogram}
\end{figure}

\begin{figure}[!h]
\centering
\includegraphics[width=\textwidth]{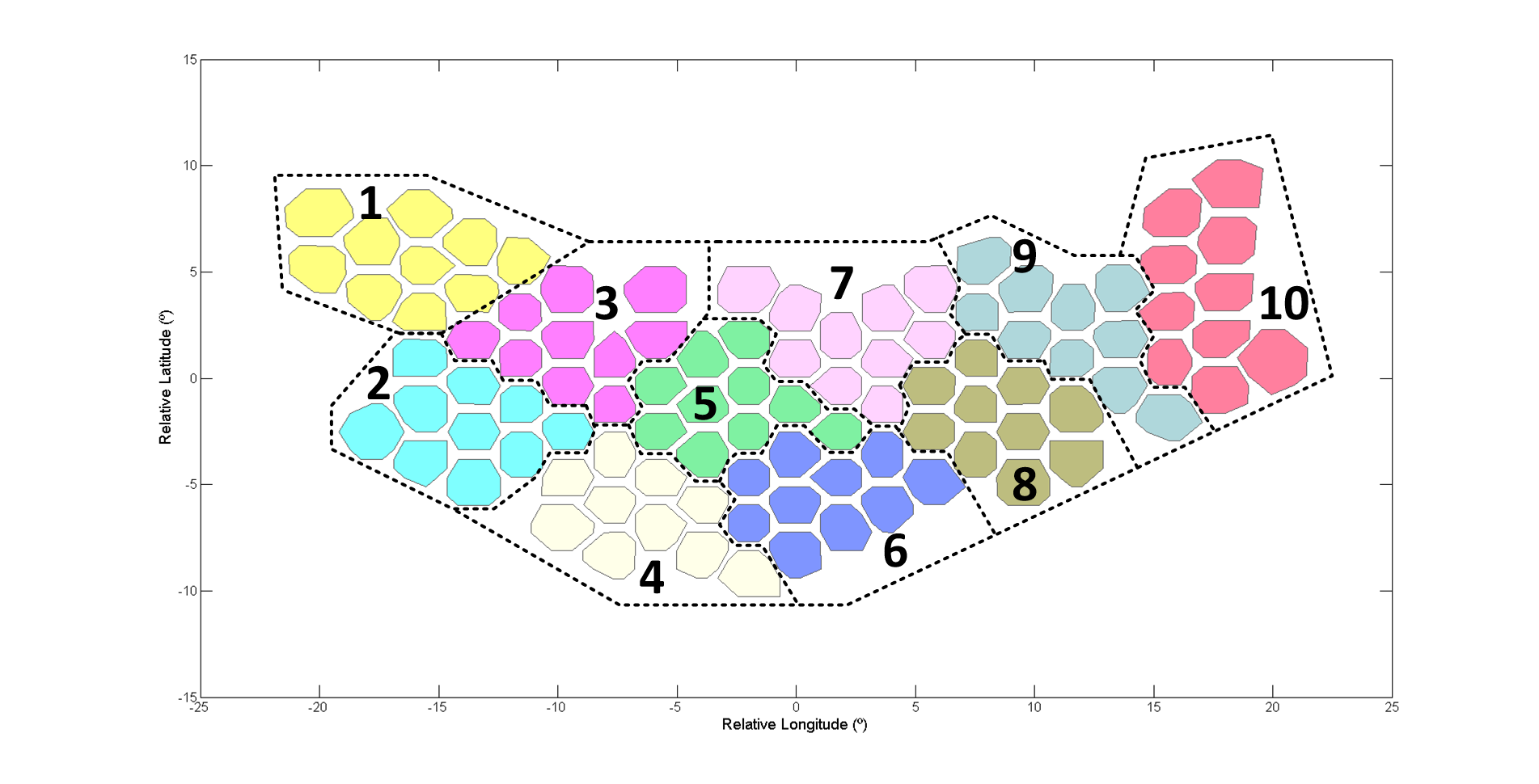}
\caption{Spot-beams are grouped into clusters.}
\label{fig:EUclusters}
\end{figure}

We have  tested the performance of the different schemes in a Monte Carlo simulation for the specifications of a multibeam satellite 
antenna which uses  a fed reflector antenna array with $N=155$ feeds to exchange signals with the users. In particular, we 
tested adaptive OBBF (labeled as OBBF-adaptive),  OBBF with coarse BFN (labeled as OBBF-coarse)  and OBBF with pre-fixed matrix (labeled as 
OBBF-pre-fixed),  with the OGBF solution in Section \ref{sec:OGBF} as reference. 
As representative 
example we have chosen the radiation pattern  provided by the European Space Agency (ESA) and used in different projects and publications by researchers cooperating in Europe with ESA, 
see, e.g., \cite{DevillersNeiraMosquera} and \cite{Vahid2016}.   This radiation pattern is designed to 
limit the level  of interference among users in systems with conservative frequency reuse and a single gateway. 
As opposed to this, we  assume that the whole  available bandwidth is used by all beams, 
resulting in high intra-cluster and inter-cluster interference levels. 
For the BFN provided by ESA, Figure \ref{fig:SIRhistogram} 
shows the histogram  of the signal to interference ratio (SIR) without precoding for full-frequency reuse, obtained from evaluating the interference for the different users 
and 100 realizations. For each realization the channel response to 100 users randomly located, one per cluster, is generated. 
As expected, many users suffer from high interference, especially those which happen to be near the edge of the corresponding beam, 
given that this BFN is suited for a unique gateway and 
 low co-channel interference associated to a  conservative frequency reuse across beams. 
In the setting under study,  the feeder link is shared by $M=10$ gateways, with the corresponding clusters shown in Figure \ref{fig:EUclusters}. 
Clusters are groups of ten beams ($k=10$). Each gateway uses only a subset of $n$ feeds, which is chosen 
by maximizing the average gain for all users in the cluster. 
The allocated power to all clusters is the same, $P_m = P/M$, with $P$ the satellite available 
transmit power. 
We assume that the different feeder links are
 transparent, neglecting the possible impairments in the communication between the gateways and the satellite. 
%By abstracting all the frequency changes and routing taking place on-board the satellite, we can consider that the design parameters are the on-board beamforming weights, preferably  fixed due to implementation constraints.  
The randomness of the Monte Carlo simulation comes from  the location of the users at the $K=100$ spot-beams; 
these locations are chosen from independent  uniform distributions inside the different beams,  with 100 users being served at each realization, 
and  independently across realizations.  \\ 

In order to compare the performance of different schemes the operation point needs to be calibrated. This is set by defining the signal to noise ratio (SNR) as 
\begin{equation}
\snr = \ex{\trace\{\H \F \F^H \H^H\}}/K
\end{equation}
and $\F$ the transmit beamforming matrix $\F = \frac{\sqrt{P}}{\sqrt{\trace\{\H^H \H\}}} \H^H$.\\

\begin{figure*}[!h]
\centering
\begin{subfigure}{.45\textwidth}
\begin{centering}
\includegraphics[width=\textwidth]{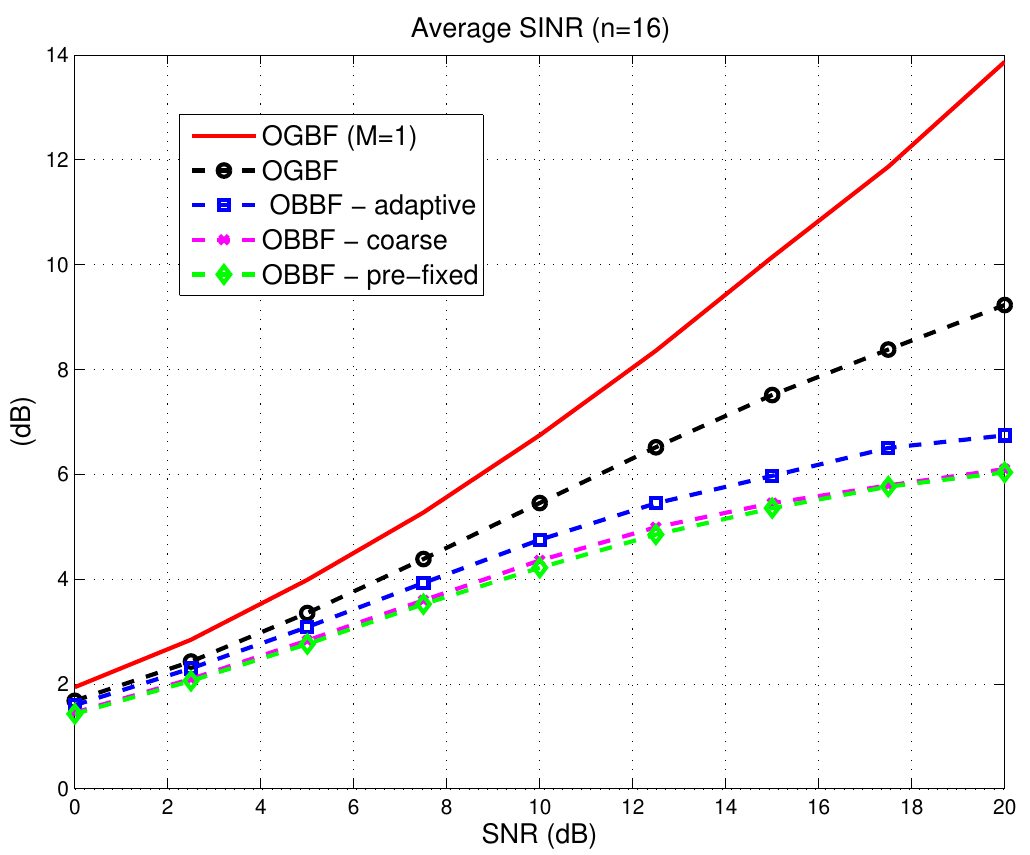}
\caption{$n=16$.}
\end{centering}
\end{subfigure}
\begin{subfigure}{.45\textwidth}
\centering
\includegraphics[width=\textwidth]{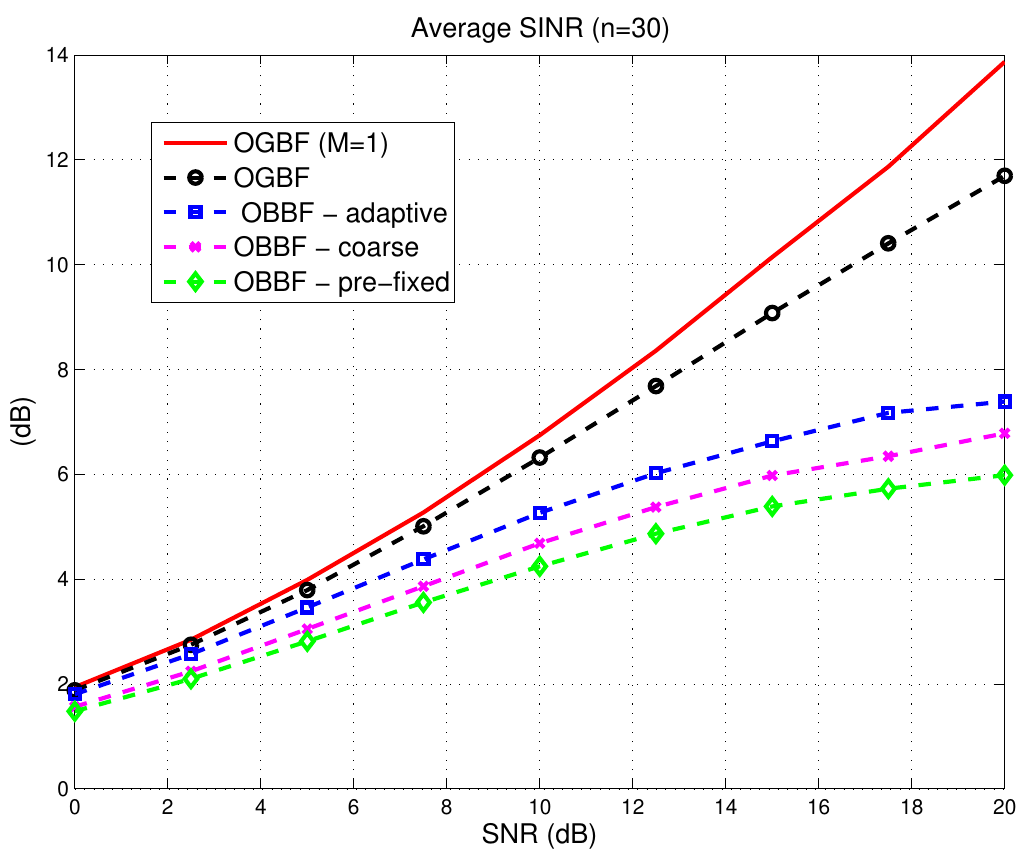}
\caption{$n=30$.}
\end{subfigure}
\caption{$M=10,k=10$, 50 realizations. Upper bound: OGBF, one gateway. Lower bound: fixed BFN provided by ESA, distributed precoders. }
\label{fig:satsimulations}
\end{figure*}

\begin{figure*}[!h]
\centering
\begin{subfigure}{.45\textwidth}
\begin{centering}
\includegraphics[width=\textwidth]{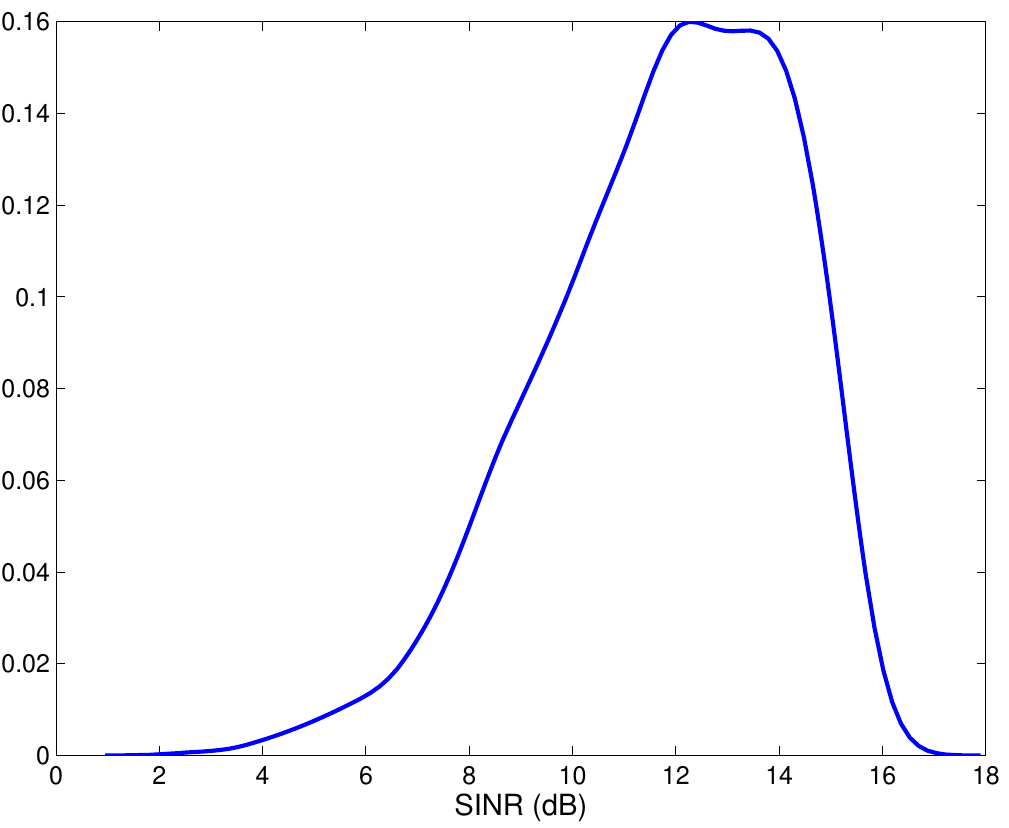}
\caption{OGBF.}
\end{centering}
\end{subfigure}
\begin{subfigure}{.45\textwidth}
\centering
\includegraphics[width=\textwidth]{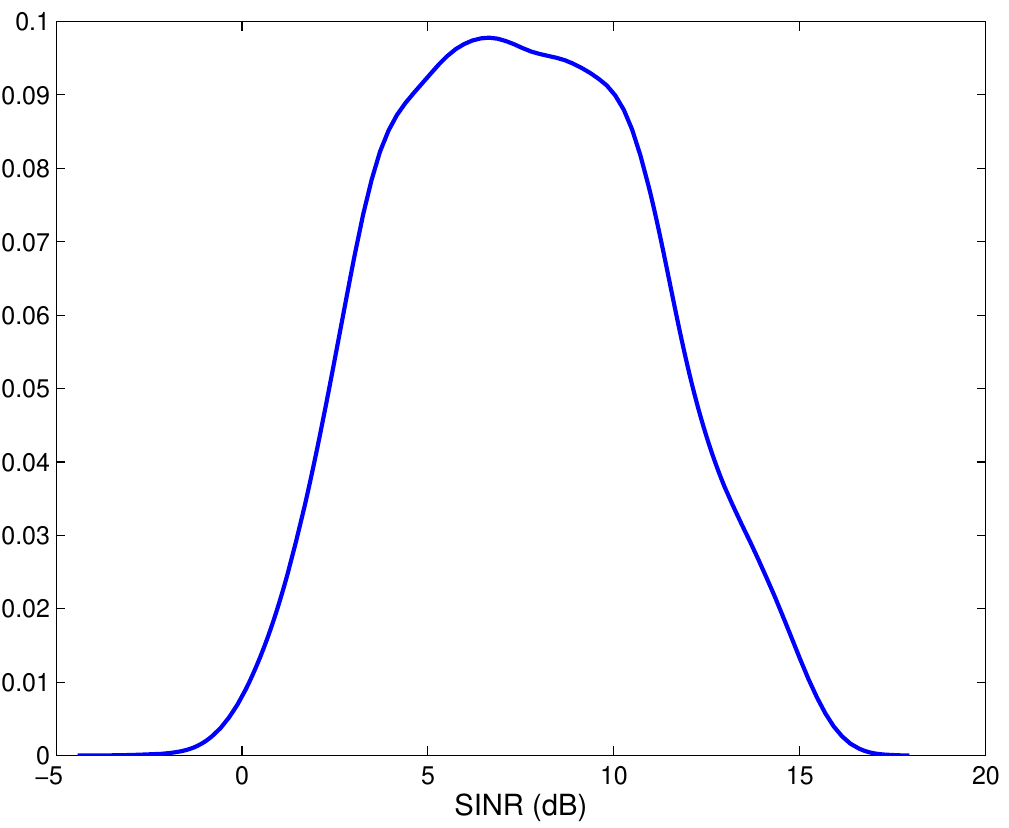}
\caption{OBBF.}
\end{subfigure}
\caption{$M=10,k=10, n = 30$, SINR histograms based on 50 realizations, SNR = 20dB. }
\label{fig:histograms}
\end{figure*}

%The design of the on-board BFN becomes instrumental, with 
Figure \ref{fig:satsimulations} presents  the average SINR for all users and the different schemes after 50 Monte Carlo realizations. 
Results have been obtained for two different number of feeds, $n=16$ and $n=30$. 
Even though all feeds are available to serve any cluster, not more than 35 participate in the 
provision of power to  an arbitrary cluster. 
The performance is upper bounded by the OGBF scheme, applicable if the gateways have access to the different feeds without an intermediate BFN and the CSIT is perfect. Even further, the OGBF bound for the single gateway case, i.e., $M=1$ and different scaling parameters for all users, is also included to illustrate the performance loss due to the lack of 
data exchange among gateways. All the other curves use an on-board BFN, 
either adaptive or fixed; the latter uses either  the coarse design  in Section \ref{sec:robust} or the BFN provided by ESA for the four-color reuse 
scheme. As expected, performance improves if more feeds are assigned to each gateway, keeping in mind that feeds can be shared by different gateways. 
There is a significant loss from the OBBF-adaptive with respect to the OGBF scheme, which increases with the SNR, and which comes from the isolated operation of the gateways. Nevertheless, the degradation of the coarse BFN  with respect to the corresponding adaptive BFN version stays below 1dB in all cases, and decreases for low SNR. It is left for future studies whether robustness can be preserved for alternative designs of the BFN able to  address the aggregated inter-cluster interference, rather than the noise or the interference to a specific set of users.  It can be also noticed that the design of a specific BFN fixed matrix as part of a global multi-gateway interference cancellation scheme  provides a gain with respect to a BFN not specifically designed with this in mind, for $n=30$, whereas for $n=16$ this gain is barely noticeable. 
Additionally, the histogram of SINR is also shown in Figure \ref{fig:histograms} for both OBBF-adaptive and OGBF schemes, with $n=30$ and 
SNR = 20dB. 
With respect to the baseline histogram in Figure \ref{fig:SIRhistogram},  a more compact distribution of SINR values is obtained. \\

\begin{figure*}[!h]
\centering
\begin{subfigure}{.45\textwidth}
\begin{centering}
\includegraphics[width=\textwidth]{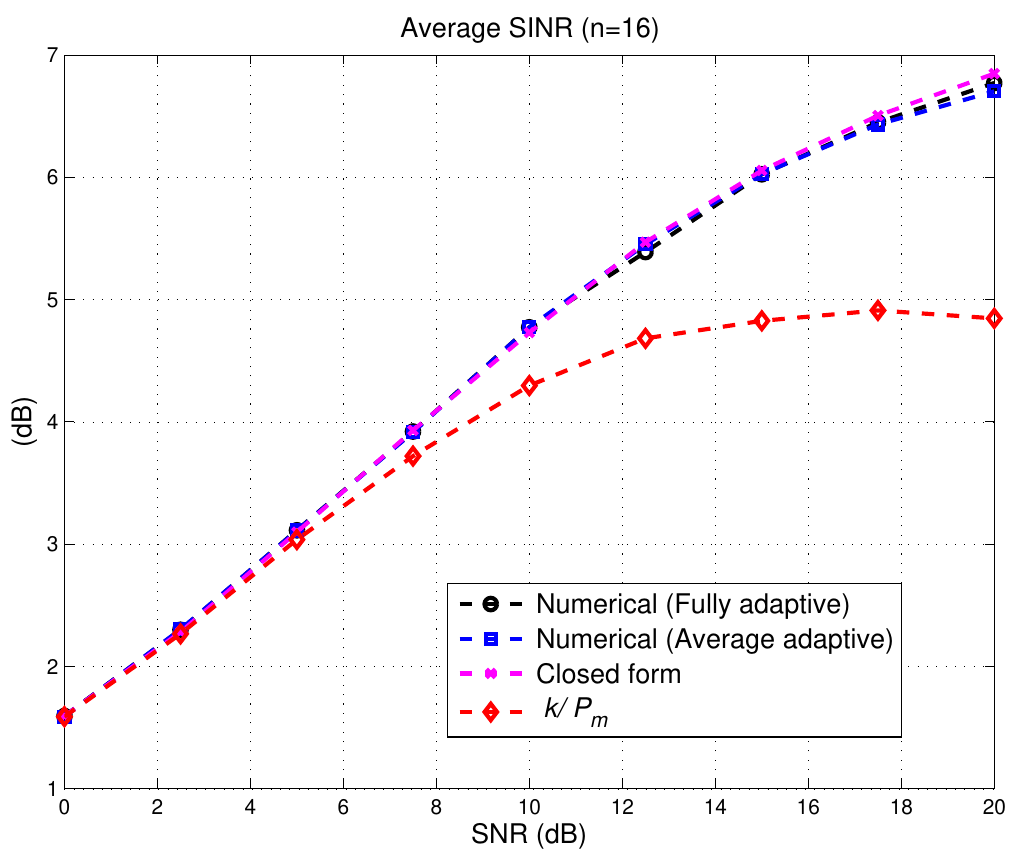}
\caption{$n=16$.}
\end{centering}
\end{subfigure}
\begin{subfigure}{.45\textwidth}
\centering
\includegraphics[width=\textwidth]{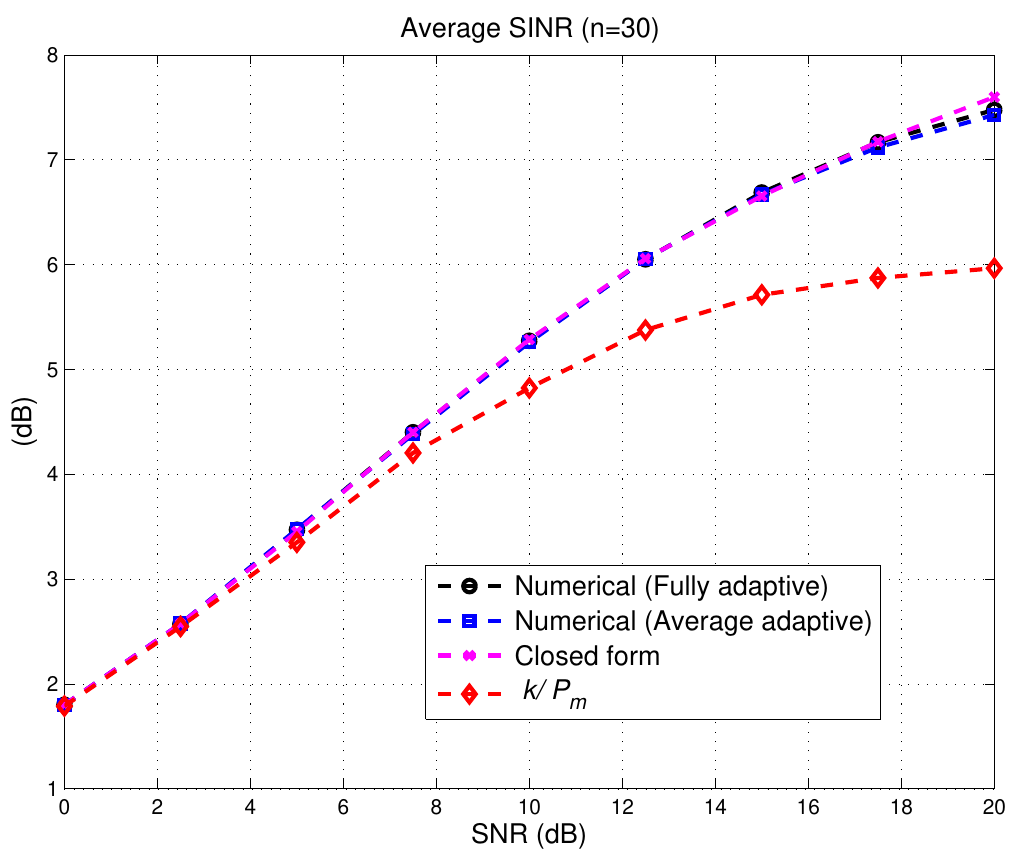}
\caption{$n=30$.}
\end{subfigure}
\caption{$M=10,k=10$, 50 realizations. Performance with different regularization factors.}
\label{fig:RegFactor}
\end{figure*}

As illustration of the  role played by the regularization factor in the precoding process, we have also compared the use of different regularization 
factors in the computation of the precoder for the adaptive BFN case (OBBF-adaptive):
\begin{enumerate}
\item $\gamma_m = k/P_m$. This is the regularization factor minimizing the MSE for a single cluster, as it is well-established in the literature 
\cite{Hochwald2005a}.
\item $\gamma_m$ the numerical solution of \eqref{eq:gamma-fixed}, with $\sigma_{ii}^{(m)}$ the $i$th diagonal entry of 
$\U_m^H\B_m^H \bm \Sigma_m \B_m \U_m$, and $\bm \Sigma_m$ approximated as \eqref{eq:approx}.
\item $\gamma_m$ the numerical solution of \eqref{eq:gamma-fixed}, with $\sigma_{ii}^{(m)}$ the $i$th diagonal entry of 
$\U_m^H\B_m^H \hat{\bm \Sigma}_m \B_m \U_m$. This 
is the regularization factor which has been used to obtain results in Figure \ref{fig:satsimulations}, which does not require inter-gateway cooperation, since $\hat{\bm \Sigma}_m$ in \eqref{eq:approx1} is based on average statistics.
\item Closed-form expression \eqref{eq:approx-gamma} to approximate the numerical solution of \eqref{eq:gamma-fixed}.
\end{enumerate}
The expectation in the last two cases has been approximated empirically. 
As depicted in Figure \ref{fig:RegFactor}, which shows the impact of the regularization factor, the inter-cluster solution $k/P_m$ falls short of being 
effective in the presence of intra-cluster interference. More interestingly, the closed form ad-hoc solution matches the performance of the 
numerically optimized regularization factor. The use of instantaneous  channel matrices does not improve significantly  the performance, and 
gives additional merit to the autonomous operation of gateways without permanent exchange of information; only average inter-cluster 
channel matrices are needed for the implementation of the proposed schemes. 
% which relies on the knowledge of the inter-cluster channels. This is due to the simplification \eqref{eq:approx1}, which affects 
%to the parameters $\sigma_{ii}$ in \eqref{eq:gamma-fixed}, which are no longer exactly the diagonal elements of $\B_m^H \bm \Sigma_m \B_m$. 

\begin{figure*}[!h]
\centering
\begin{subfigure}{.45\textwidth}
\begin{centering}
\includegraphics[width=\textwidth]{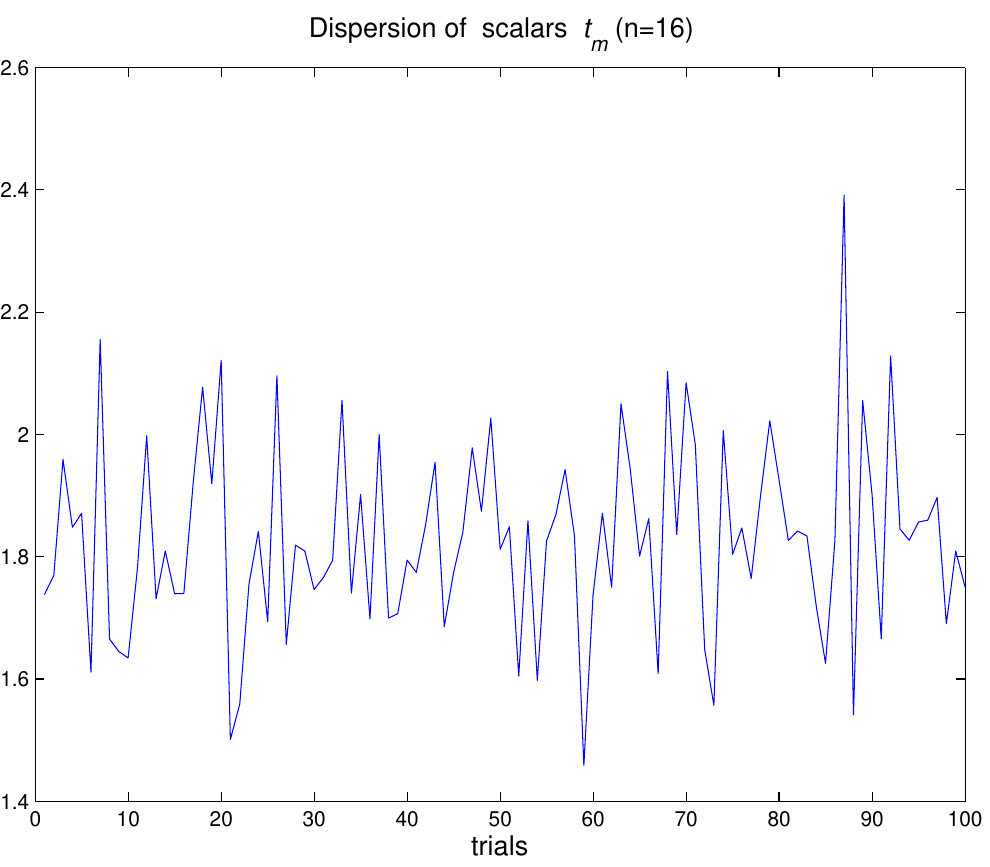}
\caption{$n=16$.}
\end{centering}
\end{subfigure}
\begin{subfigure}{.45\textwidth}
\centering
\includegraphics[width=\textwidth]{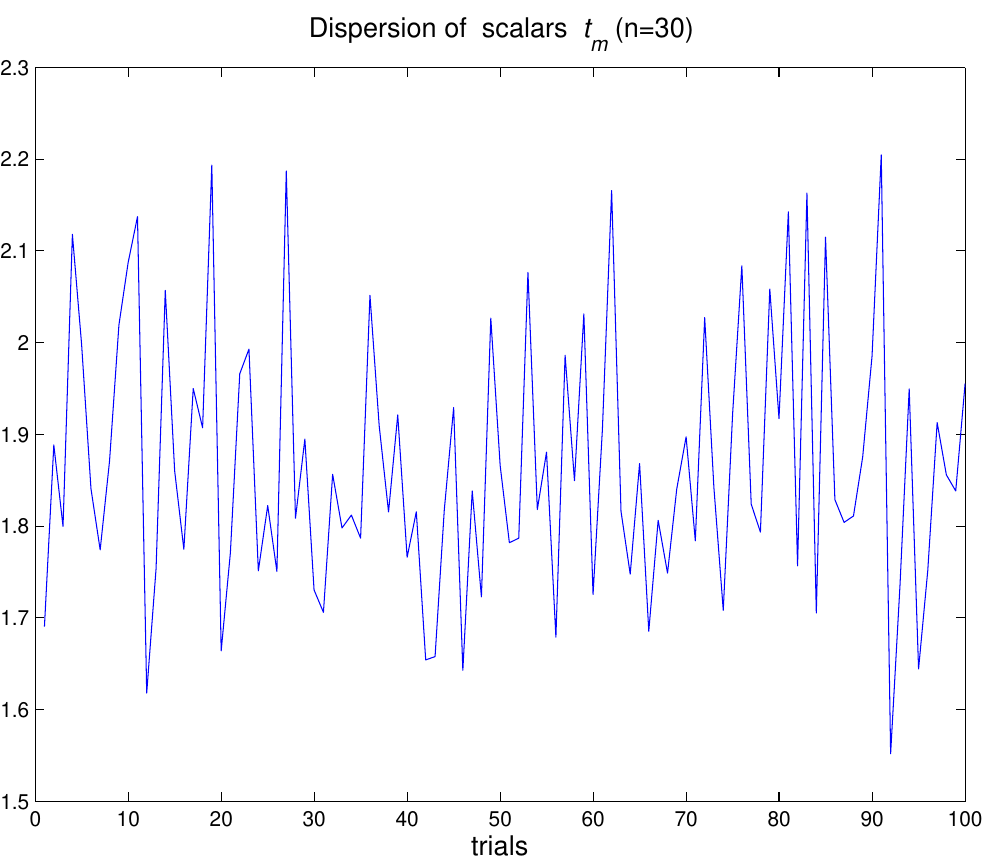}
\caption{$n=30$.}
\end{subfigure}
\caption{$\max_m\{t_m\}/\min_m\{t_m\}$ for different trials. $M=10,k=10$, SNR = 10dB.}
\label{fig:tm}
\end{figure*}

Lastly, we have checked the dispersion of the scalars $\{t_m\}$. In order to avoid the interaction among clusters, we have assumed throughout 
the paper that their values are similar. Otherwise  an iterative process to solve the multiple dependencies among 
$\{\T_m,\gamma_m,\bm \Sigma_m,t_m\}$ would require the sequential exchange of information among the gateways, and make their autonomous optimization unfeasible. 
Figure \ref{fig:tm} shows an average ratio of maximum to minimum values lower than 2 for the two settings addressed in the previous simulations, supporting the allocation of similar weights to all the inter-cluster Gramians in \eqref{eq:approx}. 

\section{Conclusions}

The mitigation of co-channel interference in multibeam satellite settings has been addressed in this paper,  for the case of 
 several ground stations using  the satellite to relay their signals to their respective clusters of beams. Both sources of interference, 
intra-cluster and inter-cluster, are attenuated by considering the global MSE and deriving distributed on-ground linear precoders and on-board 
beamforming weights. Practical rules have been provided to design the on-ground precoders and on-board beamformer for different degrees of flexibility of the latter, including also a coarse fixed BFN. Under the premise of no cooperation among 
gateways, the regularization factor of the MMSE precoders was obtained, first numerically and then in an approximated closed form, to account for 
the presence of inter-cluster leakage, generalizing existing results for a centralized precoder. 
For the purpose of benchmarking an On-Groud Beamforming solution has been derived with full flexibility.
Exploration of non-linear schemes of the type Tomlinson-Harashima to improve the rate, as proposed in 
\cite{Hochwald2005b} in combination with regularization at the transmit precoder, can be a topic for further improvement of the results exposed in this paper, together with the consideration of CSI errors in the design of the ground precoders. The design of the BFN can be also extended to account for the 
aggregated inter-cluster leakage; the key performance indicator will be the ability to mitigate the interference with a fixed design, given that a well-performing adaptive solution does not  necessarily lead to a valid robust design. 

\section*{Acknowledgement}

The authors wish to thank  Tom\'as Ram\'{\i}rez, from the University of Vigo,  for his assistance with the multi-cluster channel characterization.

\appendices 

\section{Numerical finding of Lagrange multiplier for OGBF design}

Consider the problem
\begin{equation}\label{eq:numerical-OGBF}
\min_{\F} \trace\{ \F^H \A \F - \F^H \X - \X^H \F \} \quad \text{subject to} \quad \trace\{ \F^H \F^H \} \leq P,
\end{equation}
with $\A \in \mathbb{C}^{n \times n}$  Hermitian positive definite, and $\F,\X \in \mathbb{C}^{n \times k}$.

Let $\A = \U \bm \Gamma \U^H$ be the 
eigenvalue decomposition  of $\A$, and define $\tilde \F= \U^H \F, \tilde \X = \U^H \X$. Since $\U \in \mathbb{C}^{n \times n}$ is unitary, we can reformulate 
\eqref{eq:numerical-OGBF} as 
\begin{equation}
\min_{\F} \trace\{ \tilde \F^H \bm \Gamma \tilde \F - \tilde \F^H \tilde \X - \tilde \X^H \tilde \F \} \quad \text{subject to} \quad 
\trace\{ \tilde \F^H \tilde \F^H \} \leq P.
\end{equation}
Assuming that the unconstrained solution $ \tilde \F = \bm \Gamma^{-1} \tilde \X$  is not feasible (which must be checked), then
the constraint must be satisfied with equality with $\tilde \F = (\bm \Gamma + \nu \I_n)^{-1} \tilde \X$:
\begin{equation}
\trace \{\tilde \F^H \tilde \F \} = \trace \{ \tilde \X^H (\bm \Gamma + \nu \I_n)^{-2} \tilde \X\} = P.
\end{equation}
Let $\bm \Gamma = \text{diag} \{ \begin{array}{cccc} \gamma_1 & \gamma_2 & \cdots & \gamma_n \}\end{array}$, and write 
$\tilde \X$ row-wise as 
\begin{equation}
\tilde \X = \left[\begin{array}{c} \tilde \x_1^H \\ \tilde \x_2^H \\ \vdots \\ \tilde \x_n^H \end{array}  \right].
\end{equation}
Then 
\begin{equation}
\trace \{ \tilde \X^H(\bm \Gamma + \nu \I_n)^{-2} \tilde \X \} = \sum_{i=1}^n \frac{\tilde \x_i^H \tilde \x_i}{(\gamma_i + \nu)^2} = \phi(\nu).
\end{equation}
Thus, the Lagrange multiplier must satisfy $\phi(\nu) = P$. Note that $\phi(0) > P$ (because otherwise the
unconstrained solution would be feasible) and that $\phi(\nu)$  is monotone decreasing for $\nu >0$, with
$\lim_{\nu \rightarrow \infty} \phi(\nu) = 0$; therefore there is a unique positive solution of $\phi(\nu) = P$, which can be found by
bisection, Newton's method, or any root-finding technique.

\section{Derivation of regularization factor for on-ground precoders}

We start with the decomposition $\B_m^H \H_{mm}^H \H_{mm} \B_m = \U_m \S_m \U_m^H$, which for the adaptive BFN in Section \ref{sec:adaptiveBFN} 
boils down to  $\S_{H,m}$ in \eqref{eq:decomposeHH}. The error term \eqref{eq:Em} is expressed  as 
\begin{multline}
\trace\{\E_m\} = \tr\{\I_k\} - 2 \tr\{ (\U_m\S_m\U_m^H + \gamma_m \I_k )^{-1} \U_m\S_m\U_m^H\} \\ 
+ \tr\{(\U_m\S_m\U_m^H + \gamma_m \I_k)^{-1} (\U_m\S_m\U_m^H + \B_m \bm \Sigma_m \B_m) 
(\U_m\S_m\U_m^H  + \gamma_m \I_k)^{-1} \U_m\S_m\U_m^H \} \\
+ \tr\left\{ \frac{1}{t_m} \I_k\right\}
\end{multline}
which, by using the orthonormality of $\U_m$, can be alternatively expressed as 
\begin{multline}
\trace\{\E_m\} = k - 2 \trace\left\{(\S_m + \gamma_m \I_k)^{-1}\S_m \right\} + \trace\left\{(\S_m + \gamma_m \I_k)^{-1} \cdot \right. \\
\left. (\S_m+ \U_m^H \B_m^H \bm \Sigma_m \B_m \U_m)(\S_m + \gamma_m \I_k)^{-1} \S_m \right\} + k/t_m.
\end{multline}
The scaling $t_m$ in the last term  is obtained from the power constraint in 
\eqref{eq:fixedBFN} and the precoder expression \eqref{eq:Tm}:
\begin{equation}
t_m = \frac{P_m}{\trace\left\{(\B_m^H \H_{mm}^H \H_{mm} \B_m + \gamma_m \I_k)^{-1} \B_m^H \H_{mm}^H \H_{mm} \B_m 
(\B_m^H \H_{mm}^H \H_{mm} \B_m + \gamma_m \I_k)^{-1}  \right \}}, 
\end{equation}
or, equivalently, 
\begin{equation}
t_m = \frac{P_m}{\trace\left\{ (\S_m + \gamma_m \I_k)^{-1} \S_m 
(\S_m + \gamma_m \I_k)^{-1}\right \}}.
\end{equation}
If we insert this into $\trace \{\E_m\}$, then we have the following minimization problem:
\begin{equation}\label{eq:min-regfactor}
\gamma_m = \arg \min \sum_{i=1}^k \frac{-2 \lambda_i^{(m)}}{\lambda_i^{(m)} + \gamma_m} + \frac{(\lambda_i^{(m)})^2}{(\lambda_i^{(m)}+\gamma_m)^2} + 
\frac{\sigma_{ii}^{(m)} \lambda_i^{(m)}}{(\lambda_i^{(m)} + \gamma_m)^2} + \frac{k}{P_m} \frac{\lambda_i^{(m)} }{(\lambda_i^{(m)} + \gamma_m)^2}
\end{equation}
where  $\sigma_{ii}^{(m)}$ is  the $i$th diagonal entry of $\U_m^H\B_m^H \bm \Sigma_m \B_m \U_m$, and $\lambda_i^{(m)}$ the $i$th eigenvalue of 
$\B_m^H \H_{mm}^H \H_{mm} \B_m$. By equating to zero the derivative the relation to satisfy is 
\begin{equation}\label{eq:num-gammam}
\sum_{i=1}^k \frac{\lambda_i^{(m)}}{(\lambda_i^{(m)}+\gamma_m)^3}\left(\gamma_m - \sigma_{ii}^{(m)} - \frac{k}{P_m}\right) = 0
\end{equation}
and the scaling parameter 
\begin{equation}\label{eq:tm-gammam}
t_m = P_m/\sum_{i=1}^k \frac{\lambda_i^{(m)}}{(\lambda_i^{(m)} + \gamma_m)^2}.
\end{equation}

%%%%%%%%%%%%%%%%%%%%%%%%%%%%

\bibliographystyle{IEEEtran}
\bibliography{CoarseBF}

%\bibliographystyle{unsrt}
%\begin{thebibliography}{99}

%\end{thebibliography}

\end{document}

%% file: user_cmds.tex
\def\A{{\mathbf A}}
\def\B{{\mathbf B}}
\def\C{{\mathbf C}}
\def\D{{\mathbf D}}
\def\E{{\mathbf E}}
\def\F{{\mathbf F}}
\def\G{{\mathbf G}}
\def\H{{\mathbf H}}
\def\I{{\mathbf I}}

\def\Q{{\mathbf Q}}
\def\R{{\mathbf R}}
\def\S{{\mathbf S}}
\def\T{{\mathbf T}}
\def\U{{\mathbf U}}
\def\V{{\mathbf V}}

\def\X{{\mathbf X}}

\def\Z{{\mathbf Z}}

\def\0{{\mathbf 0}}
\def\1{{\mathbf 1}}

\def\n{{\mathbf n}}

\def\p{{\mathbf p}}

\def\s{{\mathbf s}}

\def\x{{\mathbf x}}
\def\y{{\mathbf y}}

\DeclareMathOperator{\trace}{tr}

\newcommand{\openone}{\leavevmode\hbox{\small1\normalsize\kern-.33em1}}

\newcommand{\snr}{\mathop{\rm SNR}}

\newcommand{\ex}[1]{\mathbb{E}\left[ {#1}\right]}

\newtheorem{lemma}{Lemma}

%% file: Submitted_onecolumn.bbl
\begin{thebibliography}{10}
\providecommand{\url}[1]{#1}
\csname url@samestyle\endcsname
\providecommand{\newblock}{\relax}
\providecommand{\bibinfo}[2]{#2}
\providecommand{\BIBentrySTDinterwordspacing}{\spaceskip=0pt\relax}
\providecommand{\BIBentryALTinterwordstretchfactor}{4}
\providecommand{\BIBentryALTinterwordspacing}{\spaceskip=\fontdimen2\font plus
\BIBentryALTinterwordstretchfactor\fontdimen3\font minus
  \fontdimen4\font\relax}
\providecommand{\BIBforeignlanguage}[2]{{%
\expandafter\ifx\csname l@#1\endcsname\relax
\typeout{** WARNING: IEEEtran.bst: No hyphenation pattern has been}%
\typeout{** loaded for the language `#1'. Using the pattern for}%
\typeout{** the default language instead.}%
\else
\language=\csname l@#1\endcsname
\fi
#2}}
\providecommand{\BIBdecl}{\relax}
\BIBdecl

\bibitem{Minoli2015}
D.~Minoli, \emph{Innovations in Satellite Communications Technology}.\hskip 1em
  plus 0.5em minus 0.4em\relax John Wiley \& Sons, Inc., Hoboken, NJ, USA,
  2015.

\bibitem{Tronc2014}
J.~Tronc, P.~Angeletti, N.~Song, M.~Haardt, J.~Arendt, and G.~Gallinaro,
  ``Overview and comparison of on-ground and on-board beamforming techniques in
  mobile satellite service applications,'' \emph{International Journal of
  Satellite Communications and Networking}, vol.~32, no.~4, pp. 291--308, 2014.

\bibitem{Zheng2012}
G.~Zheng, S.~Chatzinotas, and B.~Ottersten, ``Multi-gateway cooperation in
  multibeam satellite systems,'' in \emph{2012 IEEE 23rd International
  Symposium on Personal, Indoor and Mobile Radio Communications-(PIMRC)}.\hskip
  1em plus 0.5em minus 0.4em\relax IEEE, 2012, pp. 1360--1364.

\bibitem{Hochwald2005a}
C.~B. Peel, B.~M. Hochwald, and A.~L. Swindlehurst, ``A vector-perturbation
  technique for near-capacity multiantenna multiuser communication-part {I}:
  channel inversion and regularization,'' \emph{IEEE Transactions on
  Communications}, vol.~53, no.~1, pp. 195--202, Jan 2005.

\bibitem{DevillersNeiraMosquera}
B.~Devillers, A.~P{\'e}rez-Neira, and C.~Mosquera, ``Joint linear precoding and
  beamforming for the forward link of multi-beam broadband satellite systems,''
  in \emph{Global Telecommunications Conference (GLOBECOM 2011), 2011 IEEE}.

\bibitem{GaudenziBook}
S.~Chatzinotas, B.~Ottersten, and R.~De~Gaudenzi, \emph{Cooperative and
  cognitive satellite systems}.\hskip 1em plus 0.5em minus 0.4em\relax Academic
  Press, 2015.

\bibitem{Vahid2016}
V.~Joroughi, M.~A. V{\'a}zquez, and A.~I. P{\'e}rez-Neira, ``Precoding in
  multigateway multibeam satellite systems,'' \emph{IEEE Transactions on
  Wireless Communications}, vol.~15, no.~7, pp. 4944--4956, July 2016.

\bibitem{Sayed2007}
M.~Sadek, A.~Tarighat, and A.~H. Sayed, ``A leakage-based precoding scheme for
  downlink multi-user {MIMO} channels,'' \emph{IEEE Transactions on Wireless
  Communications}, vol.~6, no.~5, pp. 1711--1721, May 2007.

\bibitem{Doufexi2012}
P.~Patcharamaneepakorn, S.~Armour, and A.~Doufexi, ``On the equivalence between
  {SLNR} and {MMSE} precoding schemes with single-antenna receivers,''
  \emph{IEEE Communications Letters}, vol.~16, no.~7, pp. 1034--1037, July
  2012.

\bibitem{Vandendorpe2010}
B.~K. Chalise and L.~Vandendorpe, ``Optimization of {MIMO} relays for
  multipoint-to-multipoint communications: Nonrobust and robust designs,''
  \emph{IEEE Transactions on Signal Processing}, vol.~58, no.~12, pp.
  6355--6368, 2010.

\bibitem{Stoica2004}
P.~Stoica and Y.~Sel{\'e}n, ``Cyclic minimizers, majorization techniques, and
  the expectation-maximization algorithm: a refresher,'' \emph{IEEE Signal
  Processing Magazine}, vol.~21, no.~1, pp. 112--114, 2004.

\bibitem{Ting2014}
C.~K. Wen, J.~C. Chen, K.~K. Wong, and P.~Ting, ``Message passing algorithm for
  distributed downlink regularized zero-forcing beamforming with cooperative
  base stations,'' \emph{IEEE Transactions on Wireless Communications},
  vol.~13, no.~5, pp. 2920--2930, May 2014.

\bibitem{Mosquera17submitted}
C.~Mosquera, R.~L{\'o}pez-Valcarce, and T.~Ram{\'{\i}}rez, ``Distributed
  precoding systems in multi-gateway multibeam satellites,'' 2017, submitted to
  14th International Symposium on Wireless Communication Systems.

\bibitem{Horn13}
C.~R.~J. R.~A.~Horn, \emph{Matrix Analysis}, 2nd~ed.\hskip 1em plus 0.5em minus
  0.4em\relax Cambridge University Press, 2013.

\bibitem{Devillers2011}
B.~Devillers and A.~P{\'e}rez-Neira, ``Advanced interference mitigation
  techniques for the forward link of multi-beam broadband satellite systems,''
  in \emph{2011 Conference Record of the Forty Fifth Asilomar Conference on
  Signals, Systems and Computers (ASILOMAR)}.\hskip 1em plus 0.5em minus
  0.4em\relax IEEE, 2011, pp. 1810--1814.

\bibitem{Klein2009}
Y.~C. Silva and A.~Klein, ``Linear transmit beamforming techniques for the
  multigroup multicast scenario,'' \emph{IEEE Transactions on Vehicular
  Technology}, vol.~58, no.~8, pp. 4353--4367, 2009.

\bibitem{Haardt2004}
Q.~H. Spencer, A.~L. Swindlehurst, and M.~Haardt, ``Zero-forcing methods for
  downlink spatial multiplexing in multiuser {MIMO} channels,'' \emph{IEEE
  Transactions on Signal Processing}, vol.~52, no.~2, pp. 461--471, 2004.

\bibitem{Hochwald2005b}
B.~M. Hochwald, C.~B. Peel, and A.~L. Swindlehurst, ``A vector-perturbation
  technique for near-capacity multiantenna multiuser communication-part {II}:
  perturbation,'' \emph{IEEE Transactions on Communications}, vol.~53, no.~3,
  pp. 537--544, March 2005.

\end{thebibliography}
